\documentclass[fleqn,usenatbib]{mnras}

\usepackage{newtxtext,newtxmath}

\usepackage[T1]{fontenc}

\DeclareRobustCommand{\VAN}[3]{#2}
\let\VANthebibliography\thebibliography
\def\thebibliography{\DeclareRobustCommand{\VAN}[3]{##3}\VANthebibliography}

\usepackage{graphicx}	%
\usepackage{hyperref}
\usepackage{amsmath}	%
\usepackage{float}
\usepackage{xcolor, soul}
\usepackage{booktabs} %
\usepackage[section]{placeins}
\usepackage{cleveref}

\crefname{figure}{Figure}{Figures}
\Crefname{figure}{Figure}{Figures}
\crefname{section}{Section}{Sections}
\Crefname{section}{Section}{Sections}

\usepackage{fix-cm}
\usepackage{subcaption} 
\usepackage{soul}

\usepackage{fancyhdr}  

\makeatletter
\def\ps@titlepage{
  \let\@mkboth\@gobbletwo
  \def\@oddhead{\footnotesize\@journal}

  \def\@oddfoot{
  \footnotesize\itshape
  \parbox{\textwidth}{
    © 2025 Franc O. All rights reserved. This work is made available via \textit{arXiv.org} for reading only.\\
    No part of this work may be reused, reproduced, distributed, or modified, commercially or non-commercially, without the author’s explicit written permission.}
}
  \def\@evenhead{\footnotesize\@journal\hfill} 
  \def\@evenfoot{\hfil\footnotesize\copyright\ \@pubyear\ The Authors}
  \def\sectionmark##1{}
  \def\subsectionmark##1{}
}

\title{%
Parameter Inference of Black Hole Images using Deep Learning in Visibility Space
}

\author[F. O et al.]{
Franc O $^{1,2}$,
Pavlos Protopapas $^{3}$, 
Dominic W. Pesce $^{5,6}$,
Angelo Ricarte $^{5,6}$, 
Sheperd S. Doeleman$^{5,6}$,
\newauthor
\hspace{0.05cm} Cecilia Garraffo $^{2, 6, 7}$, Lindy Blackburn$^{5,6}$, Mauricio Santillana $^{4}$
\\
$^{1}$Khoury College of Computer Sciences, Northeastern University, Boston, MA 02115\\
$^{2}$NSF Institute for Artificial Intelligence \& Fundamental Interactions (IAIFI)\\
 $^{3}$John A. Paulson School of Engineering and Applied Science, Harvard University, Boston, MA, 02134\\
$^{4}$Department of Physics, Northeastern University, Boston, MA 02115\\
$^{5}$Black Hole Initiative at Harvard University, 20 Garden St., Cambridge MA, 02138\\
$^{6}$Center for Astrophysics $\vert$ Harvard \& Smithsonian, 60 Garden St. Cambridge MA, 02138\\
$^{7}$AstroAI at the Center for Astrophysics | Harvard \& Smithsonian, Cambridge MA, 02138 \\
}

\date{Submitted to MNRAS}

\pubyear{2025}

\begin{document}
\label{firstpage}
\pagerange{\pageref{firstpage}--\pageref{lastpage}}
\maketitle

\begin{abstract}

Using very long baseline interferometry, the Event Horizon Telescope (EHT) collaboration has resolved the shadows of two supermassive black holes. Model comparison is traditionally performed in image space, where imaging algorithms introduce uncertainties in the recovered structure. Here, we develop a deep learning framework to perform parameter inference in visibility space, directly using the data measured by the interferometer without introducing potential errors and biases from image reconstruction. First, we train and validate our framework on synthetic data derived from general relativistic magnetohydrodynamics (GRMHD) simulations that vary in magnetic field state, spin, and $R_\mathrm{high}$. Applying these models to the real data obtained during the 2017 EHT campaign, and only considering total intensity, we do not derive meaningful constraints on either of these parameters. At present, our method is limited both by theoretical uncertainties in the GRMHD simulations and variation between snapshots of the same underlying physical model. However, we demonstrate that spin and $R_\mathrm{high}$ could be recovered using this framework through continuous monitoring of our sources, which mitigates variations due to turbulence. In future work, we anticipate that including spectral or polarimetric information will greatly improve the performance of this framework.

\end{abstract}


\section{Introduction} \label{sec:intro}

Black holes are regions in spacetime where gravity is so intense that nothing—not even light—can escape, defined by the presence of an event horizon \citep{EHTm87p1}. They result from extreme spacetime curvature caused by a massive amount of matter compressed into a tiny space, leading to a singularity where classical physics breaks down \citep{Penrose1965}. Studying black holes is crucial for advancing our understanding of fundamental physics, particularly under conditions that push the limits of general relativity \citep{EHTm87p2}. Observations allow scientists to explore matter behavior under extreme gravitational forces and delve into the mysteries of spacetime singularities.

Beyond their intrinsic interest, black holes serve as natural laboratories for testing the boundaries of known physical laws in extreme environments \citep{EHTm87p1}. Observing them provides insights into gravity, spacetime, and fundamental forces that govern the universe \citep{Abbott2016}. Moreover, black holes significantly influence galaxy evolution and matter distribution in the cosmos, affecting stellar dynamics and the formation of large-scale structures \citep{Gravity2019}.

The study of black holes has evolved significantly, with advances in observational techniques and theoretical models enhancing our understanding \citep{Remillard2006}. Early research relied on indirect observations, such as X-ray emissions from black hole binaries, to infer their properties and behaviors \citep{Remillard2006}. A major breakthrough occurred with the direct detection of gravitational waves from a binary black hole merger, confirming Einstein’s predictions and inaugurating a new era in astrophysics \citep{Abbott2016}.”

The Event Horizon Telescope (EHT) has revolutionized black hole research by enabling direct imaging of their immediate environments, including the supermassive black hole at the center of M87$^\star$ \citep{EHTm87p1}. These groundbreaking observations provided unprecedented insights into black hole structures and dynamics, confirming theoretical predictions and offering new avenues for testing physics in extreme gravitational fields \citep{EHTm87p1}. By combining data from telescopes worldwide, the EHT achieved a resolution capable of capturing details near a black hole’s event horizon, illuminating accretion mechanisms and jet formation processes \citep{EHTm87p2}.

The EHT was developed to directly image the environments surrounding supermassive black holes, such as M87$^\star$, enabling the study of critical physical parameters like the spin parameter $a^*$ and the asymptotic ratio of the ion to electron temperature in weakly magnetized regions in the context of a particular model $R_{\text{high}}$ \citep{moscibrodzka2016general}\citep{EHTm87p5}. Understanding these parameters is vital for deciphering matter and radiation behavior near black holes, providing insights into accretion processes and emission mechanisms that shape their observational signatures \citep{EHTm87p5}.

Analyzing the asymmetry and spatial features of black hole shadows, which correlate with $a^*$ and $R_{\text{high}}$, enables probing fundamental black hole properties and testing general relativity in strong-field regimes \citep{EHTm87p5}. Variations in the shadow’s appearance reveal information about spacetime geometry near the event horizon, accretion dynamics, and radiation emission from surrounding plasma. By studying these characteristics, we can infer the physical conditions and complex interactions between gravity, magnetism, and radiation in extreme environments \citep{EHTm87p5}.

In this research, we present a methodology that utilizes observational data from the visibility domain as input to deep learning models to predict the parameters \( a_* \) and \( R_{\text{high}} \). This approach bypasses the image reconstruction step, enhancing prediction accuracy and computational efficiency.

We employ various techniques to analyze Fourier spectrum data from simulations and observations, processing them using autoencoders, recurrent neural networks, and multitask learning architectures. This approach allows us to first learn embedded representations of the data, then capture sequential variations to ultimately predict $a_{*}$ and $R_{\text{high}}$

Following the EHT 2017 results, most research has focused on image-based analysis. \citet{Gucht2020} used SANE (Standard And Normal Evolution) simulations to regress black hole mass and accretion rates, treating $a_{*}$ as a classification problem. \citet{Lin2020} extended this by utilizing both SANE and MAD (Magnetically Arrested Disk) simulations to predict $a_{*}$.

In contrast, our approach focuses on the visibility space, bypassing the image reconstruction step. We trained models on both SANE and MAD data and applied them to EHT 2017 observations to predict $a_{*}$ and $R_{\text{high}}$ through multitask regression, achieving better performance in predicting $a_{*}$. While \citet{Lin2021} also used observed visibilities, their goal was to classify between SANE and MAD images, whereas we aim to directly predict the physical parameters $a_{*}$ and $R_{\text{high}}$.

In this study, we present a computational pipeline for analyzing black hole observational data directly in the Fourier spectrum. This pipeline has two elements. The first one consists of numerical simulations, and the second one uses the findings of the numerical simulation to analyze real observational data from the EHT. This approach may be extrapolated to other astrophysical studies where images are typically used as an intermediate step for parameter prediction. The paper is structured as follows: We discuss the synthetic numerical data generation process and the details of the acquired EHT data in Section 2. We describe in detail the computational pipeline architecture, which includes the data pre-processing and encoding steps, model assumptions, and neural network architectures in Section 3. In Section 4, we present the model calibration and out-of-sample evaluations for the synthetic, numerically generated data experiments, and demonstrate their potential generalizability to real EHT data. In Section 6, we discuss the consequences of our findings in the context of previous efforts as well as the likely uses that our methodology may have in other applications. Finally, we provide conclusions in Section 7.

\section{Data}
This study utilizes two types of data: (i) synthetic data calculated from General Relativistic Magnetohydrodynamic (GRMHD) simulations (\autoref{subsec:GRMHD}), and (ii) the real calibrated dataset from M87* EHT 2017 observations (\autoref{subsec:eht_obs}; \citealt{EHT2017CalibratedData}). The simulated data were used for model training and validation, while real observational data from the Event Horizon Telescope (EHT) 2017 observations of M87*, published in 2019, is used exclusively for inference to estimate physical parameters in the context of these models.

\subsection{GRMHD Simulations} 
\label{subsec:GRMHD}

We use GRMHD models generated using standard EHT techniques, as summarized in \citet{Wong+2022}.  In these simulations, a magnetized torus of gas is evolved in a Kerr spacetime to produce realistic fluid snapshots that are then ray-traced to produce images.  We consider both a ``Magnetically Arrested Disk (MAD)'' simulation set from \citet{Qiu2023}, characterized by dynamically important magnetic fields \citep{Narayan2003}, as well as a ``Standard and Normal Evolution (SANE)'' simulation set from \citet{roelofs2021black},
characterized by weaker magnetic fields \citep{Narayan2012}.  Current EHT studies prefer MAD models over their SANE counterparts for both Sgr A* and M87* \citep{EHTm87p8,EHTm87p9,EHTsgrap8}.

Properties of the MAD and SANE image libraries are summarized in \autoref{tab:parameter_MAD_400} and \autoref{tab:parameter_SANE_160}, respectively.  All images are ray-traced at $230~{\rm GHz}$, the same observing frequency as the real EHT data.  Prograde models are ray-traced with a viewing angle of $163^\circ$, and retrograde models are ray-traced with a viewing angle of $17^\circ$.  This is to ensure the correct orientation of the observed brightness asymmetry with the approaching jet.  Then, to ensure that the brightness asymmetry is consistent among both prograde and retrograde models, retrograde model images are all rotated by $180^\circ$.

Of particular interest for parameter inference are $a_{*}$, the BH spin parameter, and $R_\mathrm{high}$, a parameter which modulates the ion-to-electron temperature ratio, since these two particle species are not believed to be in thermal equilibrium.  The spin parameter $a_{*}$ originates from the Kerr metric, where it varies from 0 (no angular momentum) to 1 (maximum angular momentum).  Here, we use a minus sign to denote a retrograde accretion disk, where the BH and large-scale accretion disk rotate in opposite directions.  Meanwhile, as parameterized by \citet{moscibrodzka2016general}, $R_\mathrm{high}$ modulates the electron temperature via: 
\begin{align}
    \dfrac{T_{\rm ion}}{T_{\rm electron}} = \dfrac{1}{1 + \beta^2} + R_{\rm high} \dfrac{\beta^2}{1 + \beta^2},
\end{align}
\noindent where $\beta = P_{\rm gas} / P_{\rm mag}$ is the ratio of the gas to magnetic pressure.  Increasing $R_\mathrm{high}$ will tend to shift emission from the weakly magnetized disk towards the strongly magnetized jet. This effect is more pronounced for SANEs than for MADs \citep{EHTm87p5}. 

\begin{table}
\centering
\begin{tabular}{|c|p{5cm}|}
\multicolumn{2}{c}{MAD Models}                                    \\ \hline
\multicolumn{1}{|l|}{\bf{Parameter}}       & \bf{Values}                           \\ \hline
\multicolumn{1}{|l|}{$a_{*}$}           & $-0.90, -0.70, -0.50, -0.30,$ \newline $0.00, 0.30, 0.50, 0.70, 0.90$ \\ \hline
\multicolumn{1}{|l|}{$R_{\text{high}}$} & $1, 10, 20, 40, 80, 160$    \\ \hline
\multicolumn{1}{|l|}{Number of Snapshots}           & 901             \\ \hline
\multicolumn{1}{|l|}{Time Resolution} & 100 $r_gc^{-1} = 850 \ \mathrm{hr}$ \\ \hline
\multicolumn{1}{|l|}{Field of View}   & 160 $\mu$as \\ \hline
\end{tabular}
\caption{Properties of the MAD model images used in this study.}
\label{tab:parameter_MAD_400}
\end{table}

\begin{table}
\centering
\begin{tabular}{|cc|}
\multicolumn{2}{c}{SANE Models}                                    \\ \hline
\multicolumn{1}{|l|}{\bf{Parameter}}       & \bf{Values}                           \\ \hline
\multicolumn{1}{|l|}{$a_{*}$}           & $-0.94, -0.50, 0.00, 0.50, 0.94$ \\ \hline
\multicolumn{1}{|l|}{$R_{\text{high}}$} & $1, 5, 10, 20, 40, 80, 160$    \\ \hline
\multicolumn{1}{|l|}{Number of Snapshots}           & 101             \\ \hline
\multicolumn{1}{|l|}{Time Resolution} & $10 r_{g}c^{-1} \simeq 85 \text{ hrs}$ \\ \hline
\multicolumn{1}{|l|}{Field of View}   & $80 \text{ }\mu as$ \\ \hline

\end{tabular}
\caption{Properties of the SANE model images used in this study.}
\label{tab:parameter_SANE_160}
\end{table}

\begin{figure*}
    \centering
    \includegraphics[width=\textwidth]{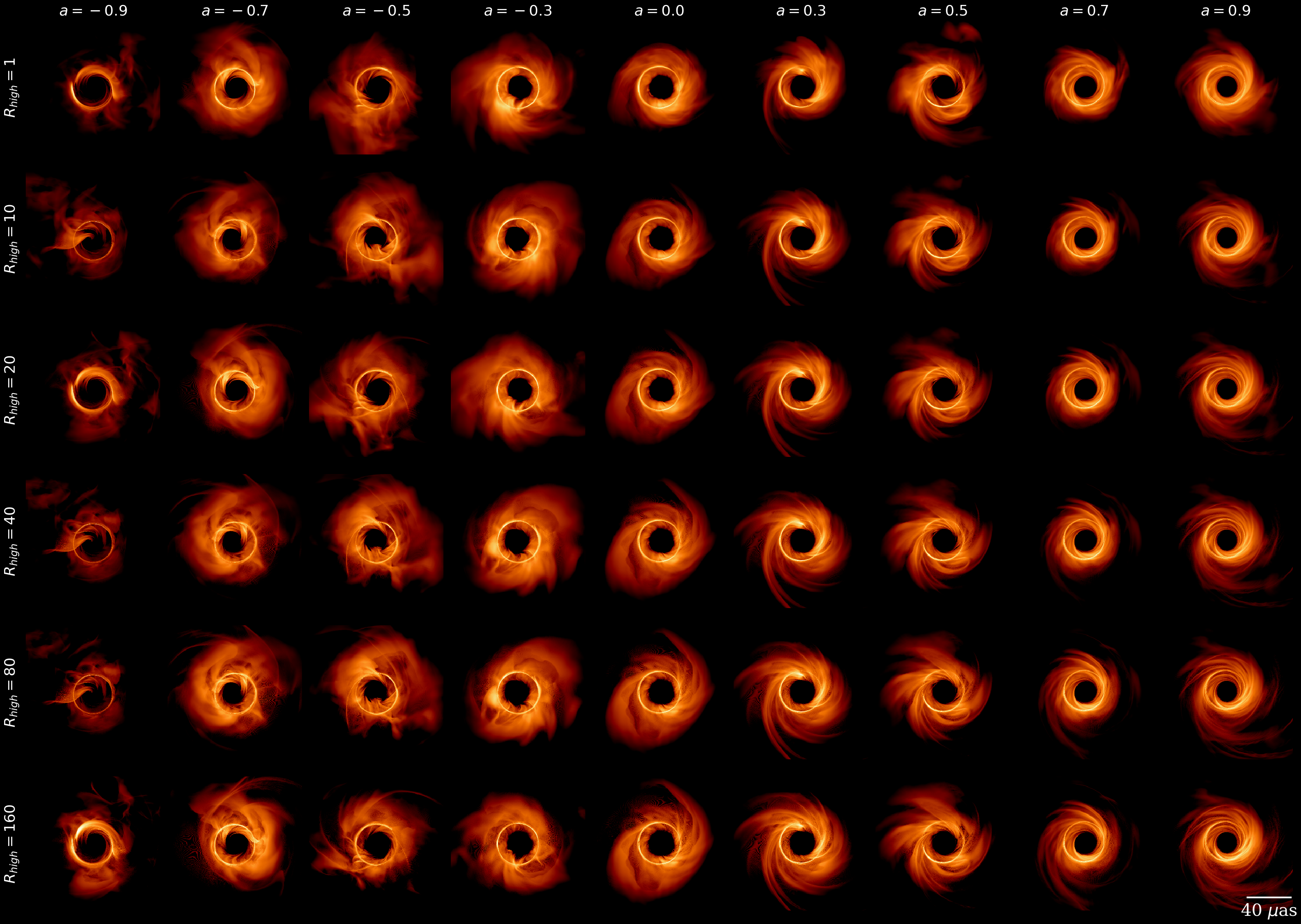}
    \caption{Gallery of example MAD snapshots used in this study.  Snapshots are plotted on a logarithmic scale with three orders of magnitude of dynamic range to better visualize low surface brightness details.  Simulations with different values of $a$ are shown in different columns, and simulations with different values of $R_\mathrm{high}$ are shown in different rows.}
    \label{fig:mad_gallery}
\end{figure*}

\begin{figure}
    \centering
    \includegraphics[width=0.47\textwidth]{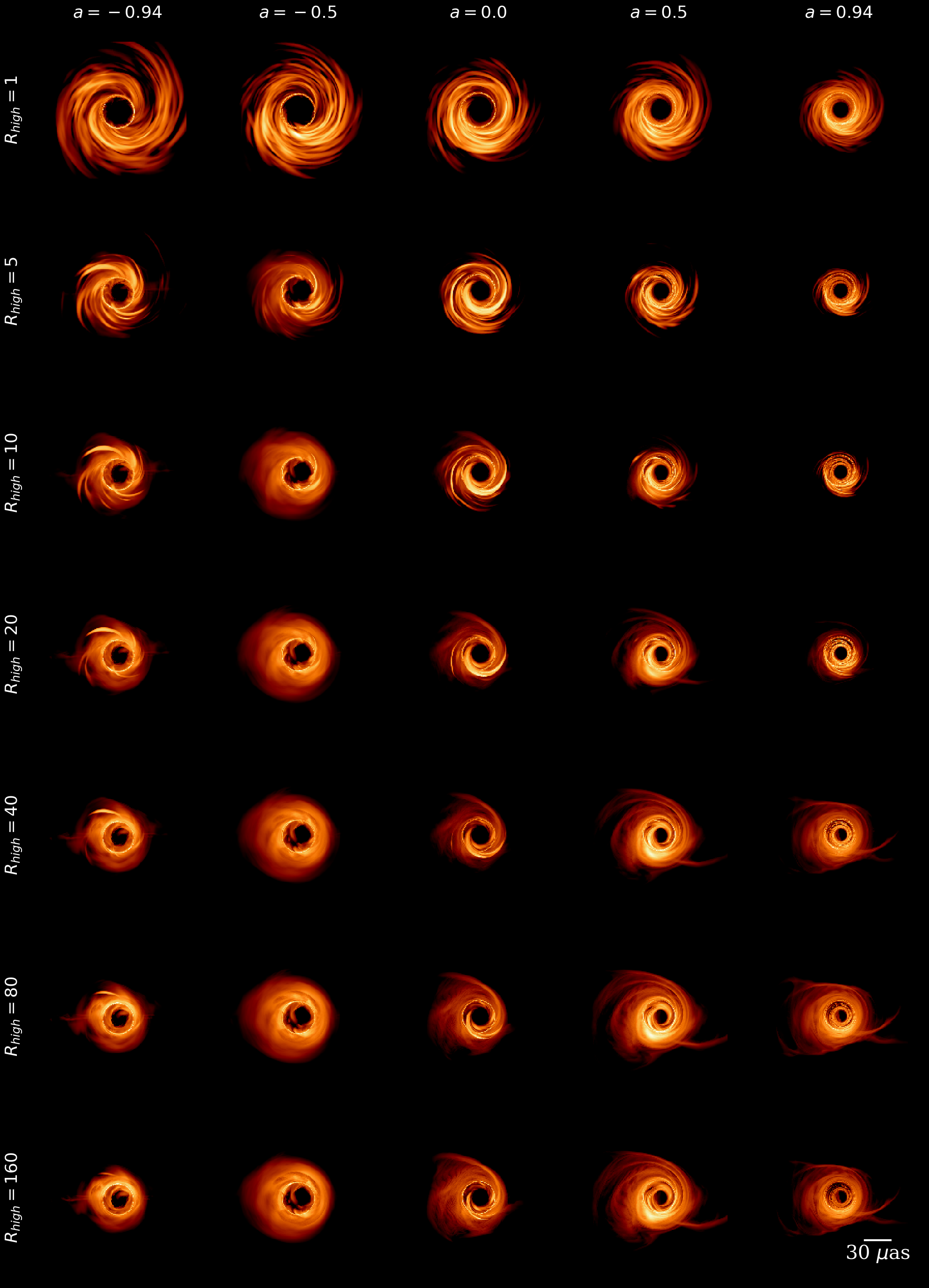}
    \caption{As \autoref{fig:mad_gallery}, but for our SANE models.  These models exhibit more obvious sensitivity to $a$ and $R_\mathrm{high}$, leading us to expect that parameter inference should be easier for SANEs than for MADs.}
    \label{fig:sane_gallery}
\end{figure}

In \autoref{fig:mad_gallery} and \autoref{fig:sane_gallery}, we visualize example snapshots from our MAD and SANE simulation libraries, respectively.  Images are plotted in a logarithmic scale with 3 decades of dynamic range to better visualize features with low surface brightness.  The photon ring, subtending 38 $\mu$as, is easily visible in each of these models.  Evolution with respect to both $a$ and $R_\mathrm{high}$ is stronger for SANEs than for MADs, leading us to expect more successful parameter inference for the SANE library. 

\subsection{Synthetic observations} 
\label{sec:synthetic_observations}

For each of the SANE and MAD images described in the previous section, we use \texttt{eht-imaging} \citep{ehtim01,ehtim02} to generate synthetic visibility datasets in a manner analogous to the procedure described in \citet{EHTm87p3}.  At a high level, this process involves Fourier transforming the image frames and sampling the complex visibilities at the same spatial frequencies --- or $(u,v)$-coverage --- as the real EHT observations.  We work with ``scan-averaged'' data, such that each complex visibility corresponds to an integration time of several minutes \citep{EHTM87Paper3}.

An interferometer like the EHT nominally samples the Fourier transform of the sky image \citep{Thompson2017}, but real observations are subject to a variety of data corruptions (``noise'') that break this ideal relationship.
We produce two different versions of each synthetic dataset
\begin{itemize}
	\item We produce a \textit{noise-free} version of each dataset, for which the complex visibilities are uncorrupted samples of the Fourier transform of the sky image.
	\item We produce a \textit{thermal-noise-only} version of each dataset, for which Gaussian-distributed thermal noise is added to each complex visibility at a level inherited from the calibrated datasets provided by the EHT Collaboration \citep{EHTM87Paper3,EHT2017CalibratedData}.
\end{itemize}

\noindent The synthetic data generation procedure described here produces $(u,v)$-coverage matching that of the real EHT observations, with noise properties that are appropriate for optimally-calibrated EHT data.

\subsection{Data Preprocessing} 
\label{sec:data_preprocessing}

Additional preprocessing steps were performed to ensure the data was suitable for training. Neural networks require input data to be normalized so that features have a consistent scale, which improves the stability and efficiency of the training process. Normalization involves transforming the data to have a mean of zero and a standard deviation of one. This was implemented using \texttt{scikit-learn}'s \texttt{StandardScaler} \citep{scikit-learn}.

To evaluate the impact of resolution on model performance, various resolutions were tested during data preprocessing. We also adjusted the dataset to various resolutions by either undersampling (binning) or oversampling (interpolating) the data. These resolutions were tested during the training and inference stages, which are described in detail in Section~\ref{subsec:GRMHD} and subsequent sections. The differences in model performance across resolutions were found to be negligible.

\subsection{EHT Observations} \label{subsec:eht_obs}

The main goal of this study is to use the trained models, developed with simulated data, to analyze real observational data of M87$^{\star}$\footnote{https://eventhorizontelescope.org/for-astronomers/data} and infer the black hole’s physical parameters. Observations from the Event Horizon Telescope (EHT) on April 5th, 6th, 10th, and 11th of 2017 provided the necessary radio-wave data for analysis.

The observational data was in uvfits format, so no further preprocessing beyond normalization with scikit-learn’s StandardScaler was necessary to prepare the visibility data for model input. Using the trained models, based on GRMHD datasets, we applied them to predict the black hole’s properties from the observations, connecting simulations with real observational data.

\section{Computational Pipeline Architecture} \label{sec:models}

The architecture used for this study is designed to optimize the prediction of black hole physical parameters, $a_{*}$ (spin) and $R_{\text{high}}$, from the u-v visibilities. The models are categorized into non-sequential (static) and sequential (dynamic) designs, as shown in Figures \ref{fig:cnn_reg_non_seq} and \ref{fig:cnn_reg}, respectively.

The static architecture is the most appropriate for current real observations, as current observational datasets primarily consist of single static frames or very limited temporal coverage. This design focuses on extracting spatial features from the visibility data without requiring temporal information, making it well-suited for analyzing existing Event Horizon Telescope (EHT) data.

The dynamic architecture, in contrast, is primarily intended as a feasibility study. It explores the potential for leveraging temporal information to improve predictions, assuming that future observations might involve multi-frame or time-series data. This design evaluates the model's ability to capture temporal dynamics, which could enhance predictions when richer observational datasets become available.

Below, we provide a high-level overview of the architecture to outline the sequence of steps. Following this, we go into the details of each component, offering references to the original work as well as standard textbooks for those interested in further reading.

\subsection{Static Architecture}
The non-sequential architecture consists of the following components:
\begin{enumerate}
    \item \textbf{Autoencoder:} 
The autoencoder reduces the dimensionality of the high-dimensional input visibility data while preserving critical features. This step enhances computational efficiency and removes noise, enabling the downstream models to focus on the most relevant data representations.
      \item \textbf{1D CNN:} The convolutional layer captures local spatial features within the reduced visibility space. These features are critical for identifying patterns in the data, such as intensity gradients and periodic signals, which correlate with $a_{*}$ and $R_{\text{high}}$.
    \item \textbf{Multitask Regressor:} This component predicts the two physical parameters, $a_{*}$ and $R_{\text{high}}$, in a multitask learning framework. Sharing representations between these tasks helps the model learn their interdependencies and improves prediction accuracy.
\end{enumerate}

\subsection{Dynamic Architecture}
The sequential architecture builds on the static framework and adds the ability to model temporal dynamics:
\begin{enumerate}
    \item \textbf{Autoencoder:} As in the static model, the autoencoder reduces dimensionality while preserving key features, serving as the foundational step in the pipeline.
    \item \textbf{Pretrained Language Model (LM) with Bi-directional LSTM:} The LM is used to capture temporal relationships within sequential data, with pretrained embeddings transferred to a Bi-LSTM during fine-tuning. This combined approach enables the model to interpret time-evolving patterns in visibility data effectively.
    \item \textbf{Multitask Regressor:} As in the static architecture, this component predicts $a_{*}$ and $R_{\text{high}}$.
\end{enumerate}

For both architectures, the components are seamlessly integrated into a unified pipeline. This pipeline is designed to optimize performance in predicting the target parameters, $\vec{\theta} = (a_*, R_{\text{high}})$, by efficiently combining feature extraction, temporal or spatial modeling, and multitask regression. Each element plays a critical role in ensuring accurate and robust predictions.

\begin{figure*}
    \centering
    \includegraphics[width=0.80\textwidth]{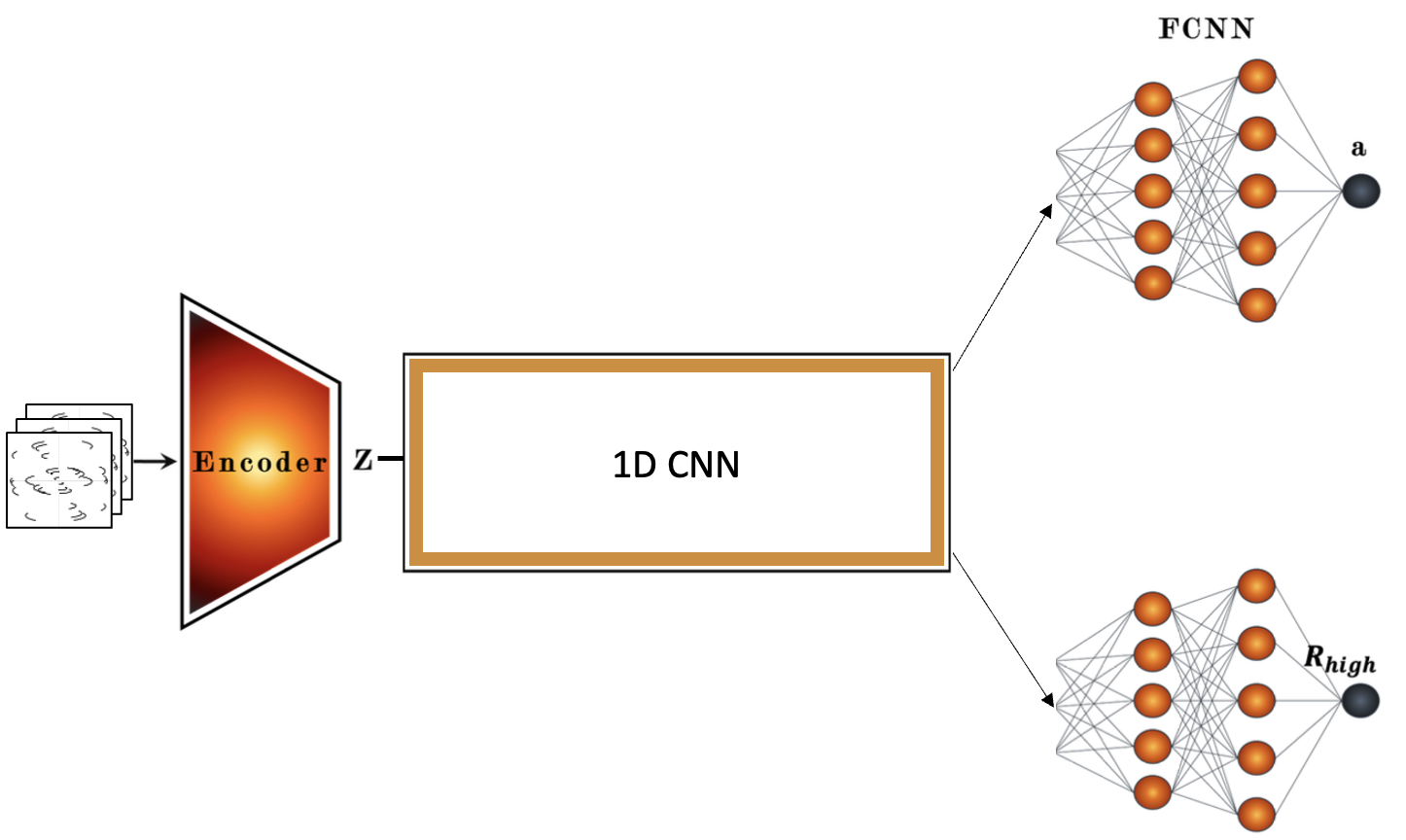}
    \caption{The architecture of the entire pipeline for non-sequential models processes radio images of shape $ B \times M \times 4$, where $ B $ represents the number of radio images in a batch, \( M \) denotes the number of frequency values in each radio image, and the four channels correspond to the \( u \), \( v \), \(\Re\) (real part), and \(\Im\) (imaginary part) components. The encoder transforms the original data into a vector \( Z \), where \( Z \in \mathbb{R}^{50} \). A 1D CNN takes \(Z\) as input and outputs a 128-dimensional vector. Finally, two fully connected neural networks (FCNNs) map this to two single output variables: \( \hat{a_*} \) or \( \hat{R}_{\text{high}} \). 
    }
    \label{fig:cnn_reg_non_seq}
\end{figure*}
\begin{figure*}
    \centering
    \includegraphics[width=0.80\textwidth]{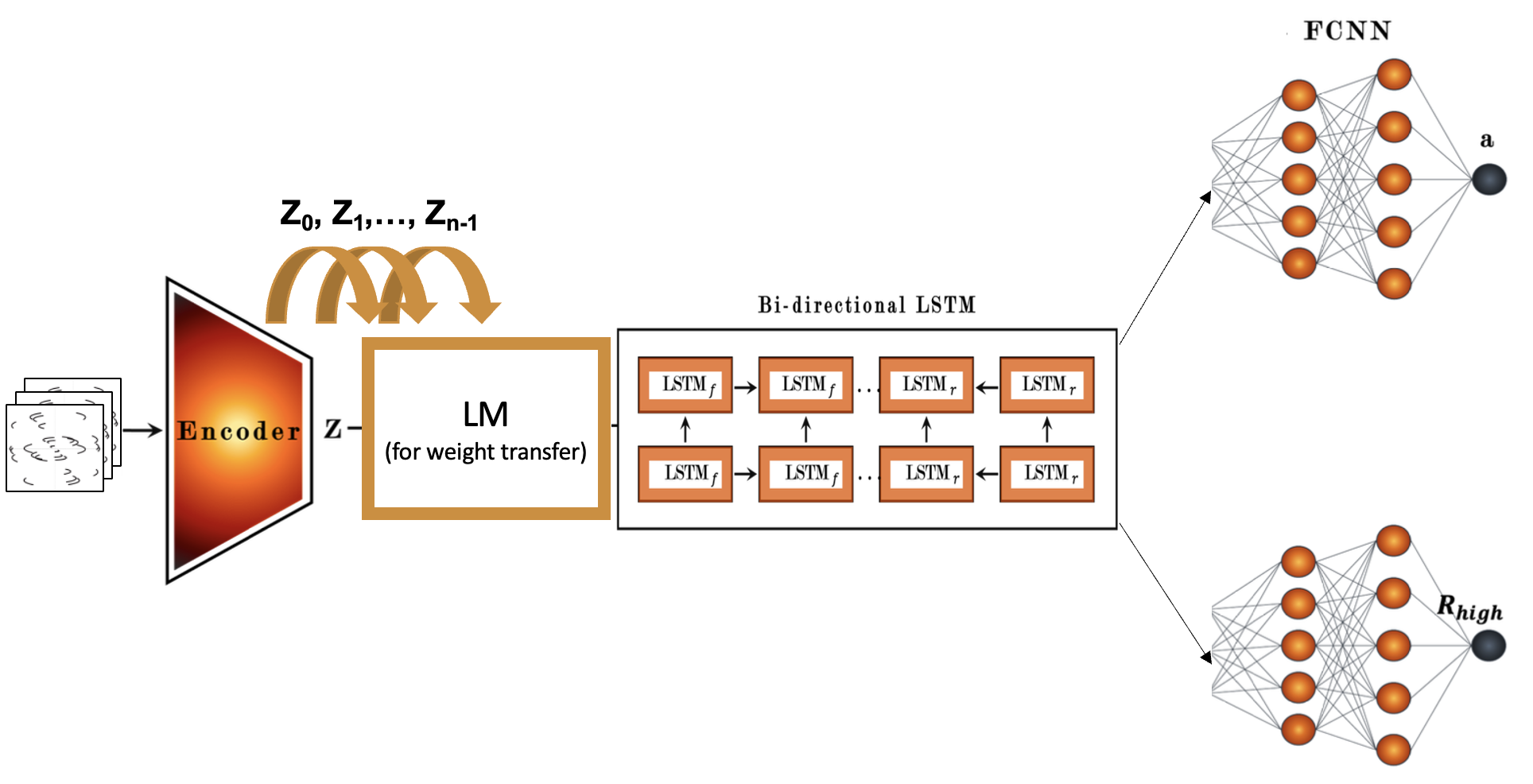}
  \caption{The architecture of the entire pipeline for sequential models processes radio images of shape \( B \times T \times M \times 4 \), where \( B \) represents the batch size, \( T \) denotes the number of frames, \( M \) corresponds to the number of frequency values in each radio image, and the four channels represent the \( u \), \( v \), \( \Re \) (real part), and \( \Im \) (imaginary part) components.  
The encoder transforms the original data into a sequence of vectors \( Z_t \), where \( Z_t \in \mathbb{R}^{50} \). The language model (LM), which is a Bi-LSTM, is trained in a self-supervised fashion using the sequence of \( Z_t \) as input, learning by predicting masked \( Z_j \) values.  
After pretraining the LM, the Bi-LSTM network trained in the LM step is fine-tuned to generate a 128-dimensional embedding, which is then fed into a fully connected neural network (FCNN) to predict \( \hat{a_*} \) and \( \hat{R}_{\text{high}} \).}
    \label{fig:cnn_reg}
\end{figure*}

\begin{figure}
    \centering
    \includegraphics[width=0.47\textwidth]{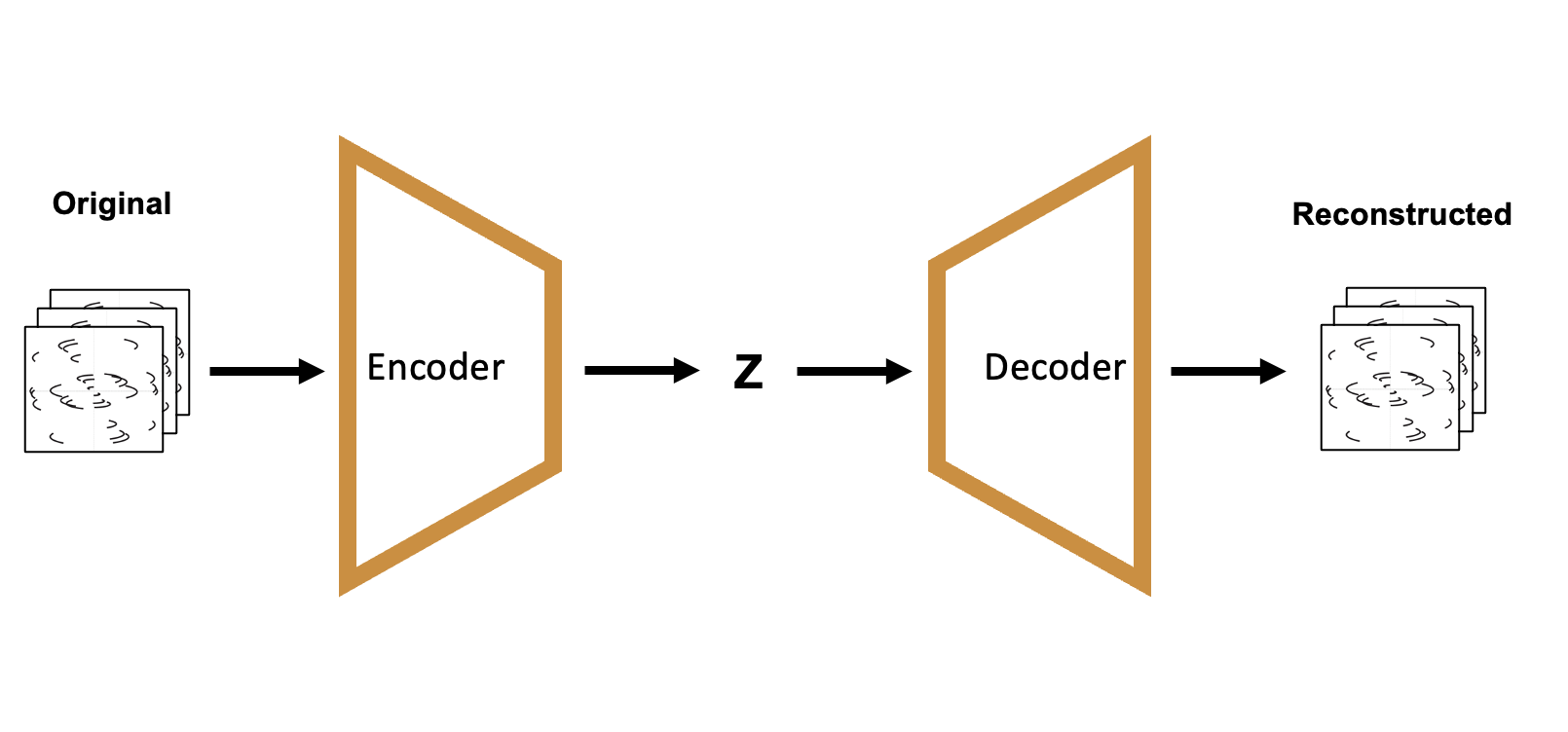}
    \caption{The architecture encodes radio images, starting with the encoder, where the original data is processed, and then moves to the decoder, which generates the reconstructed version. When the decoder is removed, the original data is transformed into a 50-dimensional vector $Z$. }
    \label{fig:ae}
\end{figure}

\subsection{Architectural Components}
In this section, we provide a more detailed description of the computational pipeline architectures as shown in Figures \ref{fig:cnn_reg_non_seq} and \ref{fig:cnn_reg}. For specifics on the parameters used, such as the number of nodes, activation functions, and other configurations, please refer to Appendix \ref{appendixA}.

\subsubsection{Autoencoder} \label{sec:Autoencoders}
Autoencoders are a type of artificial neural network designed to learn efficient codings of high-dimensional data by compressing the input into a lower-dimensional latent space and then reconstructing the original input from this compressed representation. In the context of black hole data from the Event Horizon Telescope (EHT), autoencoders can significantly reduce the dimensionality of the visibility data while preserving its critical features. This process enhances computational efficiency by removing noise and redundant information, enabling downstream models to focus on the most relevant data representations, such as key structures in the image or time-domain features. The autoencoder architecture typically consists of an encoder that maps the input to a latent space, and a decoder that reconstructs the input from this space. This unsupervised learning approach helps in extracting meaningful features from complex datasets without requiring labeled data \citep{Hinton2006}. By reducing the dimensionality in this way, autoencoders also mitigate the risk of overfitting, which is crucial when working with data that may have limited samples and a high signal-to-noise ratio \citep{Vincent2010}. This capability is especially useful for astronomical data and time-series analysis, where distinguishing subtle patterns in large datasets is key \citep{Goodfellow2016}.

This type of architecture aims to learn a representation of the data by reconstructing the original data \citep{Schmidhuber2015}.

The standard autoencoder (AE) presents three main parts: encoder, latent or embedding space (Z), and decoder.
The encoder learns the representations, storing them in the embedding space, to later use the decoder for reconstructing them (See Figure \ref{fig:ae}).

In our study, we used this architecture to learn a representation of the data, transforming it to a different size. The encoder is fed a batch of \( u \)-\( v \) data represented as \( B \times M \times 4 \) (where $ B $ represents the number of visibility images in a batch, \( M \) denotes the number of frequency values in each radio image, and the four channels correspond to the \( u \), \( v \), \(\Re\) (real part), and \(\Im\) (imaginary part) components), reduces the dimensionality to \( B \times 50 \), and then decodes it to \( B \times M \times 2 \). Both the encoder and decoder are trained to minimize the reconstruction error, but we only use the last two columns of the data, i.e., the real (\( \Re \)) and imaginary (\( \Im \)) parts of the signal, while ignoring the \( u \) and \( v \) values.  
The \textbf{loss function} used is the mean squared error (MSE) between the input real and imaginary components and their reconstructed counterparts (output of the decoder).  

After training the entire architecture, we remove the decoder and use the encoder to transform the data into a new latent space.

\subsubsection{1D CNN for Static Feature Extraction} \label{sec:1D_CNN}

Convolutional Neural Networks (CNNs) are a class of deep learning architectures designed to extract spatial patterns from structured data \citep{lecun1989backpropagation, lecun1998gradient}. Although traditional two-dimensional CNNs are widely used for image data, one-dimensional CNNs (1D CNNs) \citep{waibel1989phoneme} are better suited for processing sequential data, such as the visibility spectrum in this study.

1D CNNs utilize convolutional filters to identify localized features along a single dimension, followed by pooling layers to reduce dimensionality and emphasize key characteristics. This architecture effectively captures static or localized patterns, such as intensity gradients, edge transitions, or periodic signals, while maintaining computational efficiency \citep{kiranyaz20211d}. The hierarchical design allows for progressively abstract feature extraction as the data propagates through deeper layers. %

In this study, 1D CNNs are employed in the static model to extract static features from individual frames of GRMHD-simulated data, which have been dimensionally reduced by the encoder described above. These static features include spatial properties and intensity variations, which are crucial for understanding the physical states of black holes independent of their temporal evolution. By isolating static feature learning, the 1D CNN focuses on spatial characteristics, providing a robust framework for predicting black hole parameters from both simulated and real observational data.

\subsubsection{Language Model} \label{sec:LM}

A language model (LM) is a type of neural network designed to capture the structure and relationships in sequential data. In this study, we use a Bi-directional Long Short-Term Memory (Bi-LSTM) network inspired by ELMo \citep{peters2018elmo} as part of our \textbf{sequential model} to learn temporal patterns, such as how events evolve over time.  

The LM is pretrained in a self-supervised manner using \textbf{masked prediction}, where parts of the input sequence are hidden, and the model learns to reconstruct them. This encourages it to capture meaningful dependencies within the sequence. The \textbf{loss function} used for training is the mean squared error (MSE) between the masked and predicted values.  

For an intuitive introduction to  ELMo and contextualized word representations, readers can refer to Jay Alammar’s blog \citep{alammar2019elmo}. A broader discussion on the evolution of NLP models, including ELMo, can be found in Sebastian Ruder’s blog \footnote{\href{https://ruder.io/nlp-imagenet/}{Ruder's blog on NLP advancements}}.  

\subsubsection{Sequential Modeling with Pretrained Bi-LSTM for Intermediate Prediction} \label{sec:Bi-LSTM}

To capture temporal dependencies in the GRMHD-simulated data, we employ a  Bi-directional Long Short-Term Memory (Bi-LSTM) network, pretrained as a language model (LM) to predict intermediate inputs. This pretraining step enables the model to learn embeddings that encode temporal relationships effectively, forming a strong foundation for downstream tasks.  

The Bi-LSTM architecture \citep{schuster1997bidirectional} processes sequences in both forward and backward directions, mimicking how we understand a story, considering both past and future events. During pretraining, the Bi-LSTM learns to predict missing or intermediate steps in a sequence, akin to filling in the blanks of a story. For example, given the start and end of an event, the model infers what might happen in between. This captures crucial temporal dependencies, essential for modeling the evolution of black holes. 

The GRMHD data is structured as sequences of frames, similar to movie clips. Each frame provides static features, or snapshots of information, while the sequence encodes the system’s temporal evolution.   

Once the Bi-LSTM has learned general patterns during pretraining, we adapt it for black hole parameter regression through fine-tuning. This helps the model focus on predicting both short-term dynamics (local changes) and long-term trends (global behavior) in the data. By leveraging the pretrained embeddings, the fine-tuned Bi-LSTM significantly improves model performance on downstream tasks.

For readers unfamiliar with Bi-LSTMs, \citet{olah2015understanding} may provide an intuitive explanation.

\subsubsection{Multitask Regressor} \label{sec:Multitask}

The multitask regressor estimates black hole parameters, such as \(a_*\) and \(R_{\text{high}}\), by processing outputs extracted from either the  1D CNN in the static model or the  Bi-LSTM in the sequential model. The network consists of fully connected layers with ReLU activations, gradually narrowing to compress features and focus on high-level abstractions.  

The architecture consists of \textbf{two separate regressors}, each responsible for predicting a specific parameter. Each regressor has its own output layer, consisting of a final dense layer with a single unit and linear activation, producing continuous scalar outputs for \(a_*\) and \(R_{\text{high}}\).  

The regressors are trained using a multitask loss function that combines mean squared error (MSE) for each parameter prediction, enabling the model to leverage shared knowledge across tasks and improve generalization.  

This architecture is applied to both static and sequential data:  
\begin{itemize}
    \item For sequential data, the Bi-LSTM model processes temporal sequences, and its final embedding serves as input to the regressor.
    \item For static data, the 1D CNN extracts features from individual snapshots, which are then used by the regressor to independently predict \(a_*\) and \(R_{\text{high}}\).
\end{itemize}

The details of the architecture are provided in the appendix.

\subsection{Computational Elements}

To train the models, we used the mean squared error (MSE) as the loss function and the Adaptive Moment Estimation (Adam) algorithm \citep{kingma2014adam}. Training was primarily conducted on a single A100 GPU, resulting in training times ranging from 30 minutes to 2 hours. For more details on the training process and hyperparameter settings, please refer to Appendix~\ref{appendixA}.

\section{Results}
\subsection{GRHMD Simulation Analysis}

In this section, we analyze the results of models trained on simulations, as outlined in Section \ref{sec:models}. Our focus is on using synthetic data derived from GRMHD simulations to predict the physical parameters of M87$^{\star}$. This analysis serves as a precursor to the application of these models to real observational data, which will be discussed in Section \ref{sec:real_data}.

This study formulates the retrieval of two black hole parameters as a regression problem with continuous outputs:
\begin{enumerate}
\item \textbf{Spin} ($a_{*}$): Correlated with the asymmetric emission ring due to relativistic beaming, influenced by its magnitude ($|a_*|$) and direction (prograde or retrograde).
\item \textbf{The asymptotic ratio of the ion to electron temperature in weakly magnetized regions in the context of a particular model} ($R_{\text{high}}$): Larger values of $R_\mathrm{high}$ typically concentrates emission towards more magnetized regions, such as the jet funnel.
\end{enumerate}

Model performance is evaluated using $ R^2$ \footnote{\( R^2 \) is the proportion of explained variance over total variance.}, where values closer to one indicate better predictions.

It is important to note that, from this point onward, a logarithmic transformation was applied to  $R_{\text{high}}$  during both the training and testing phases:

$$\hat{R}_{\text{high}} = \frac{\log(R_{\text{high}})}{\log(160)} .$$
This transformation normalized the distribution of \( R_{\text{high}} \), reducing skewness and variance. By compressing the dynamic range, it mitigates the dominance of large values, preventing the model from being disproportionately influenced by high-magnitude inputs. This helps stabilize gradient updates during training, reducing the risk of vanishing or exploding gradients and improving overall convergence.

\subsubsection{Static Models for Parameter Regression}
\label{sub:param_reg_static}
We first explore static models trained on simulations where time is not considered, treating each observation independently. These models are trained on synthetic observations derived from GRMHD simulations, without explicitly modeling sequential dependencies. By ignoring temporal information, we focus on capturing the spatial structure of the system, relying solely on individual frames for inference.

\subsubsection*{Training Strategies for Static Models}

\noindent The data was split into 80\% training, 10\% validation, and 10\% test sets to ensure a balanced distribution while maintaining sufficient data for model optimization and evaluation. The training set (80\%) was used for learning model parameters. The remaining 20\% of the data was further split into validation (10\%) and test (10\%) sets using a stratified random split to preserve data distribution. The validation set was used for hyperparameter tuning and early stopping, while the test set served as an independent evaluation benchmark. The splits were performed using the \texttt{train\_test\_split} function from Scikit-learn.

\smallskip
\subsubsection*{Model Performance For Static Models}

\noindent We performed regression on $a_{*}$ and $\hat{R}_{\text{high}}$ using a static approach, employing an autoencoder and a 1D Convolutional Neural Network (CNN) combined with a multitask regressor, as described in Sec.~\ref{sec:models}.

We evaluated the static model on synthetic data generated from SANE at 230 GHz 
and MAD at 230 GHz simulations. The data included statistical errors simulated with the \texttt{ehtim} library, allowing us to assess robustness to observational noise.

The model’s performance was quantified using the $R^2$ metric under two conditions: no noise and statistical errors. All the results are shown in Tab.~\ref{tab:table_GRMHD_no_seq}.

Under ideal conditions (no noise), the static model showed strong performance for SANE data, with the  230 GHz dataset achieving the highest \( R^2 \) for \( a_* \) (92.40\%) and 
the best \( R_{\text{high}} \) performance (66.93\%). Introducing statistical errors led to a slight decline in performance. SANE 230 GHz remained quite robust, with \( R^2 \) for \( a_* \) decreasing marginally to 90.69\%.

In contrast, MAD at 230 GHz did not yield satisfactory results. %

\begin{figure*}
 
    \centering
    \begin{subfigure}{0.44\textwidth}
         \centering
        \includegraphics[width=\textwidth]{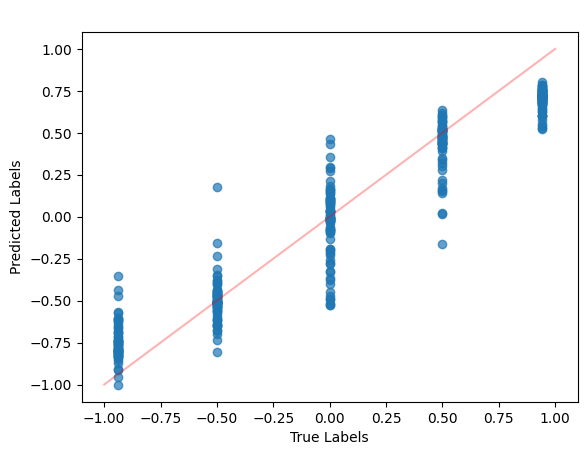}

   \end{subfigure}
   \hfill
    \begin{subfigure}{0.44\textwidth}
        \centering
        \includegraphics[width=\textwidth]{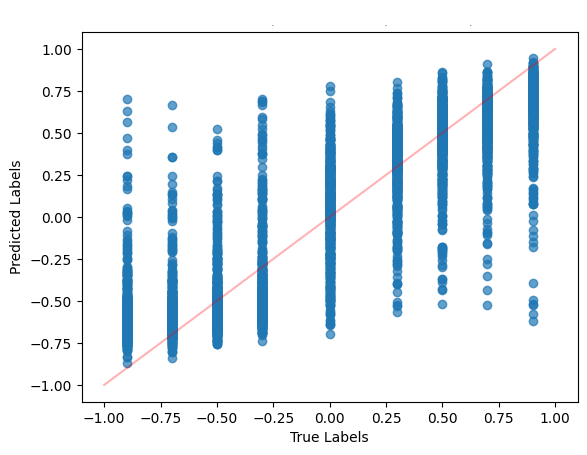}
        
    \end{subfigure}
    \caption{Scatter plot comparing true labels and predicted labels for the spin from (a) the SANE 230 GHz Static Model (no noise) on the left and (b) the MAD 230 GHz Static Model (no noise) on the right. The red line represents the ideal 1:1 relationship, indicating that predictions generally align with true values for the SANE data. However, it also illustrates that predictions for the MAD data are not well constrained, which is consistent with the $R^2$ value we obtained.\label{fig:grmhd_true_pred_SANE_MAD_static_230}
    }
\end{figure*}

While R$^2$ provides a high-level summary of model performance, a more detailed assessment requires examining the alignment between true and predicted values.  \cref{fig:grmhd_true_pred_SANE_MAD_static_230},  present scatter plots comparing predicted and true labels for the spin parameter $a_{*}$ under different conditions. The red line represents the ideal 1:1 relationship, showing that predictions for the SANE data generally align with true values. However, it also highlights that predictions for the MAD data are less constrained, which aligns with the $R^2$ value we obtained.

\begin{table}
  \caption{Results from our experiments with GRMHD Data on static models. The results are based on the test set. 
  The statistical errors are provided by the ehtim script applied to the synthetic observations.}
  \label{tab:table_GRMHD_no_seq}
  \centering
  \setlength{\tabcolsep}{3pt}
   \begin{tabular}{lcccll}
\toprule
GRMHD Data & Parameter(s) & No noise $R^2$ (\%) & Stat. error  $R^2$ (\%) \\
\midrule
  & $a_{*}$ & \textbf{92.40} & \textbf{90.69} \\
SANE 230 GHz & $R_{\text{high}}$ & 66.93 & 48.71 \\
\hline
\noalign{\vskip 1mm}
  & $a_{*}$ & \textbf{42.83} & \textbf{27.16} \\
MAD 230 GHz & $R_{\text{high}}$ & 02.07 & 00.99 \\
\bottomrule
\end{tabular}
\end{table}

\subsubsection*{Key Observations For Static Models}

\begin{enumerate}
    
    \item The MAD data showed weak performance, likely due to reduced feature distinguishability in low magnetization scenarios, which hampers accurate regression.
    
    \item The performance of the SANE 230 GHz  
    may be more robust under conditions involving statistical errors.

    \item It is easier to predict $a_{*}$ than $R_{\text{high}}$, likely due to the latter’s higher sensitivity to noise and corruptions, which complicates its estimation.
\end{enumerate}

\subsubsection{Sequential Models for Parameter Regression} \label{sec:param_reg_sequential}

We now explore models trained on simulations with time as an additional dimension, where sequential dependencies are explicitly modeled. These models are trained on synthetic observations derived from GRMHD simulations. %
By including temporal information, we aim to better capture the dynamics of the system.

\subsubsection*{Impact of Sequence Length:}
\begin{figure}
    \centering
    \includegraphics[width=0.47\textwidth]{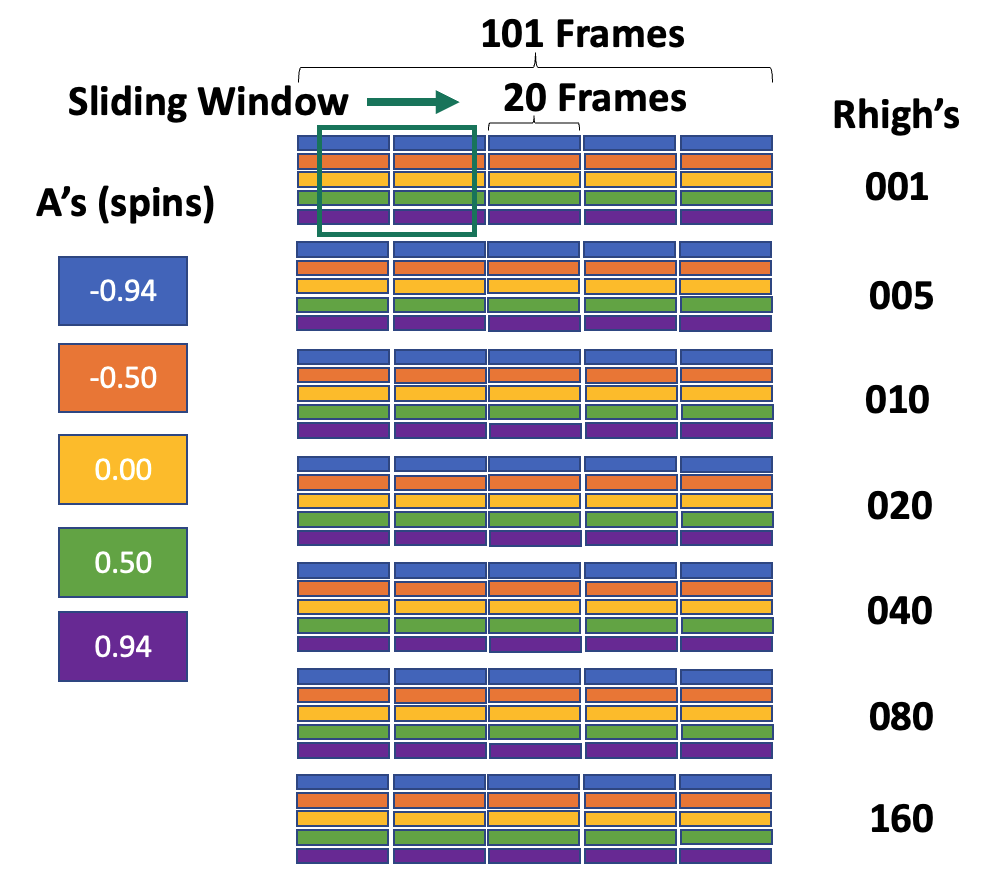}
    \caption{SANE 230 GHz example of sequence segmentation with sequential data and a window size of 20 frames.}
    \label{fig:Segments}
\end{figure}

\begin{figure*}
    \centering
    \begin{subfigure}{0.44\textwidth}
\includegraphics[width=1.0\textwidth]{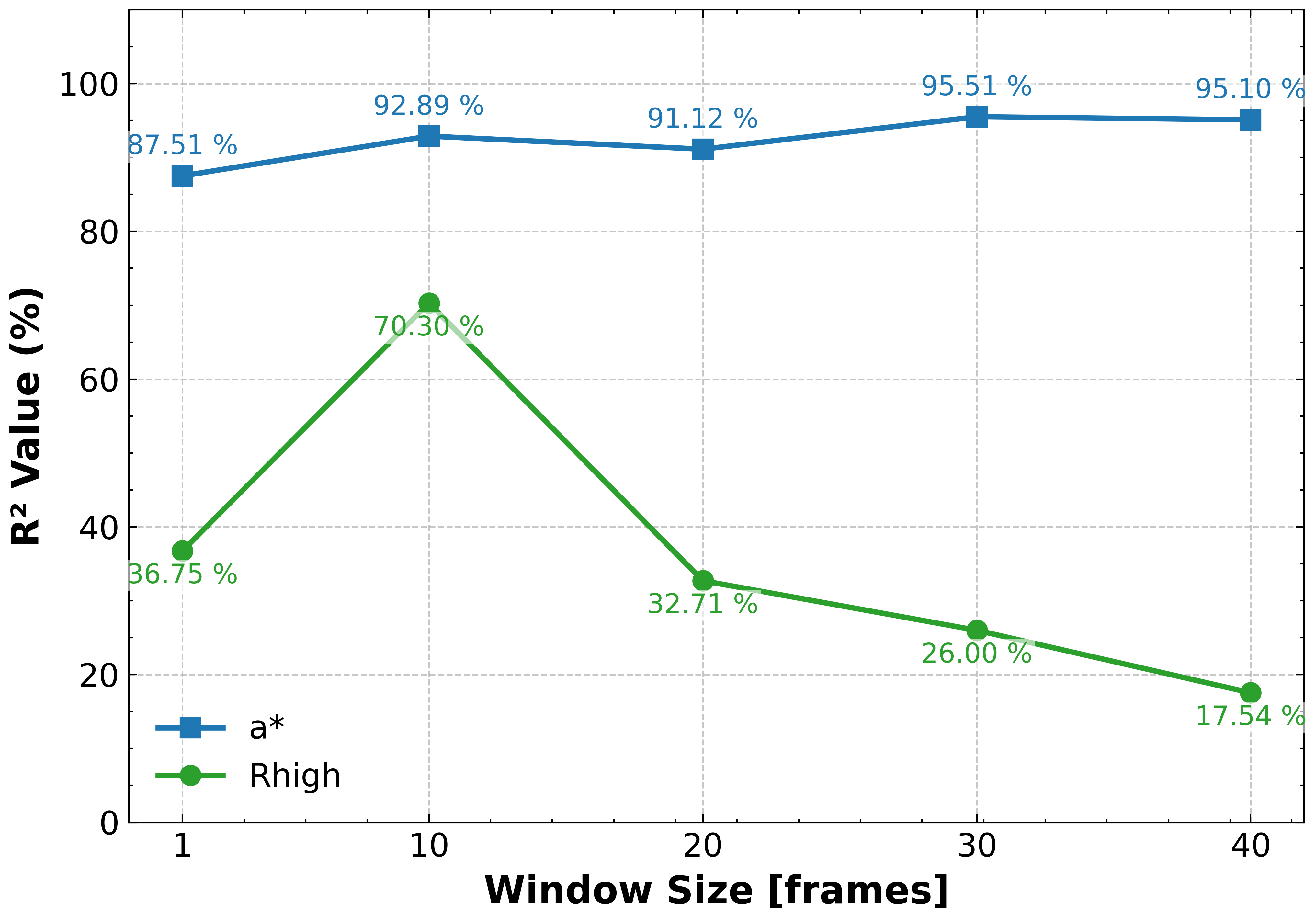}
 \end{subfigure}
 \hfill
 \begin{subfigure}{0.44\textwidth}
  \includegraphics[width=1.0\textwidth]{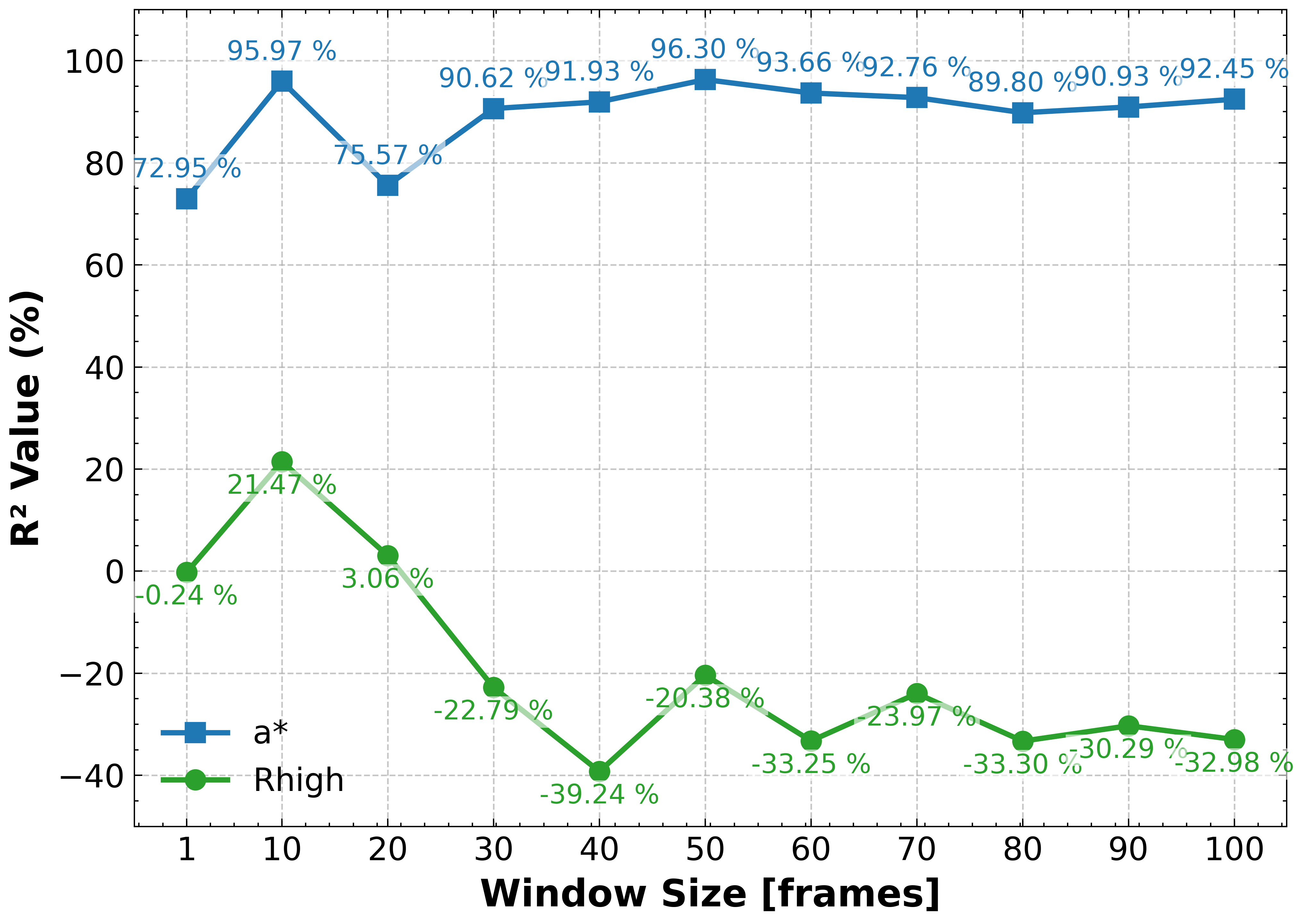}

\end{subfigure}

\caption{Comparison of $R^2$ values for the SANE 230 GHz dataset (left) and MAD 230 GHz dataset (right) as a function of window size in frames for $a_{*}$. The optimal sequence length is approximately 10 frames for SANE and between 10 and 50 frames for MAD. Notably, the time resolution in the MAD data is ten times higher than in the SANE data. \label{fig:Segments_windows_plot}}
\end{figure*}

To examine the role of sequence length, we trained models on datasets with varying numbers of frames per sequence, as shown in Figure \ref{fig:Segments}. Longer sequences provide additional temporal context, but also result in smaller training sets. This trade-off limits the amount of data available for training. \cref{fig:Segments_windows_plot} shows that while longer sequences offer more temporal context, the resulting improvements in model performance were marginal or even worse, suggesting that most of the critical information for regression is already encoded in the u-v spectrum features. The optimal sequence length was approximately 10 frames for SANE and between 10 and 50 frames for MAD. Due to memory constraints and comparable performance across different lengths, we used 10 frames for MAD.

\subsubsection*{Training Strategies for Sequential Models}
To optimize training of the sequential models, we first pre-trained the Bidirectional Long Short-Term Memory (Bi-LSTM) model as a language model using reconstruction Mean Squared Error (MSE) as the loss function. The model was trained to capture temporal dependencies in the data. Once pre-trained, we transferred the weights from this language model and fine-tuned them in conjunction with the regressor. During fine-tuning, the loss function was updated to the MSE of the true values of  $a_{*}$  and  $\hat{R}_{\text{high}}$, allowing the model to adjust its parameters for the regression task. Additionally, we applied L2 regularization uniformly across all Bi-LSTM and regression layers to mitigate overfitting and improve generalization. Distinct learning rates were assigned to the pre-trained and Bi-LSTM layers to balance stability and adaptability during training.

\smallskip
To ensure effective model training, validation, and evaluation, we employed distinct partitioning strategies for the SANE and MAD datasets, tailored to their respective structures.

\smallskip
For the \textbf{SANE dataset}, the data consists of 35 time series, covering 5 spin values and 7 \( R_{high} \) values, each containing 101 frames.  
Within each time series, 61 frames were segmented as described above using a sliding window of 10 frames and used for training. The validation set comprised 20 frames, which were segmented in the same way and used to tune the hyperparameters. Finally, the remaining 20 frames were segmented and used for testing to allow independent performance evaluation.

 For the \textbf{MAD dataset}, the partitioning strategy differed due to variations in the data structure. Here, the data contains 54 time series, covering 9 spin values and 6 $ R_{high} $ values, each containing 901 frames. Within each time series,  861 frames were assigned to the training set. The validation set consisted of 20 frames, segmented into two sets of 10 frames, which were used for hyperparameter tuning and model selection. Finally, the remaining frames served as an independent evaluation benchmark to assess model generalization.

These partitioning strategies were designed to maximize the availability of training data while preserving distinct validation and test sets, to prevent overfitting and ensure robust performance assessment.

\begin{table}
  \caption{Results from our experiments with GRMHD data on sequential models. The results are based on the test set.
  The statistical errors are provided by the ehtim script applied on the synthetic observations.}
  \label{tab:table_GRMHD_seq}
  \centering
  \setlength{\tabcolsep}{3pt}
   \begin{tabular}{llllll}
\toprule
GRMHD Data & Parameter(s) & No noise $R^2$ (\%) & Stat. errors  $R^2$ (\%)\\
\midrule
  & $a_{*}$ & \textbf{89.92} & \textbf{93.25} \\
SANE 230 GHz & $R_{\text{high}}$ & 70.51 & 60.86 \\
\hline
\noalign{\vskip 1mm}
  & $a_{*}$ & \textbf{92.54} & \textbf{91.49} \\
MAD 230 GHz & $R_{\text{high}}$ & 61.04 & 39.05 \\
\bottomrule
\end{tabular}
\end{table}
\smallskip
\subsubsection*{Model Performance for Sequential Models Overview:}
Sequential models demonstrate robust effectiveness across all models and tasks. Table~\ref{tab:table_GRMHD_seq} illustrates that the models reach $R^2$ values of up to 93.25\% for $a_{*}$ and 70.51\% for $\hat{R}_{\text{high}}$ under both noise-free and noisy conditions, indicating their capability to accurately capture underlying physical parameters.

\begin{figure*}
    \centering
    \begin{subfigure}{0.44\textwidth}
    \centering
    \includegraphics[width=\linewidth]{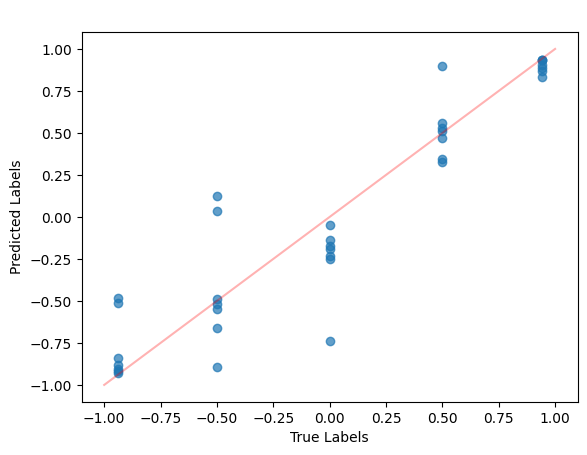}
   
     \end{subfigure}
    \hfill
  \begin{subfigure}{0.44\textwidth}
    \centering
    \includegraphics[width=\linewidth]{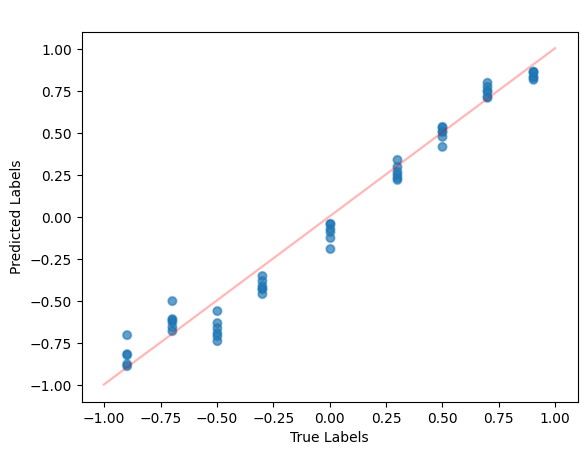}

   \end{subfigure}
\caption{Scatter plot comparing true labels and predicted labels for the spin from (a) the SANE 230 GHz Sequential Model (no noise) on the left and (b) the MAD 230 GHz Sequential Model (no noise) on the right. The red line represents the ideal 1:1 relationship, showing that predictions generally align with true values but exhibit some expected scatter.  \label{fig:grmhd_true_pred_SANEMAD_seq_230}}
\end{figure*}

It is intriguing that in this scenario, models trained on MAD data achieve comparable performance (except for $\hat{R}_{\text{high}}$ with noisy data) in the sequential setting. This indicates that the discriminating characteristics for $a_{*}$ and $\hat{R}_{\text{high}}$ in MAD data lie within the dynamics.
In \cref{fig:grmhd_true_pred_SANEMAD_seq_230}, scatter plots are presented that compare the predicted and actual labels for the spin parameter $a_{*}$ associated with two sequential models. Analysis of these plots reveals that the predictions generally follow the anticipated trend throughout the entire domain, although some expected scatter is present.

\subsubsection*{Key Observations for Sequential Models}

\begin{enumerate}
    \item Significant improvements were observed across all experiments, demonstrating the effectiveness of sequential models in capturing temporal dependencies.
    
    \item For static models trained with MAD data, performance was limited, indicating that MAD data alone may not provide enough information. However, when using sequential data, we observed a marked improvement, suggesting that incorporating temporal information is crucial for leveraging MAD data effectively.

    \item All models showed robustness to statistical errors.
    
\end{enumerate}

\subsection{Real Observations Analysis on M87\texorpdfstring{$^{\star}$}{*}  - EHT 2017}\label{sec:real_data}

In this section, we discuss the results of applying our models—created using simulated SANE and MAD GRMHD data—to deduce the black hole parameters $a_{*}$ and $R_{\text{high}}$ from the observational data obtained by EHT M87$^*$.

\smallskip
We excluded models trained on sequential data from this experiment as the actual observational data is not dynamic.
\smallskip

We used the real EHT observations presented in Section \ref{subsec:eht_obs}, consisting of single high- and low-band u-v spectrum data from April 5th, 6th, 10th, and 11th, 2017. These observations combine to form a total of eight data sets. For each of the eight observations, the u-v plane sampling is distinct. To accommodate this, eight sets of simulated u-v data were created using SANE and MAD GRMHD 230 GHz simulations, employing the same u-v sampling to ensure alignment with the observations. Subsequently, eight separate models, the AE-1D-CNN regressors, were trained for each respective dataset.

Despite exploring various model architectures and configurations, we were unable to obtain consistent predictions using MAD-trained static models. Here, \textit{consistent} predictions refer to achieving comparable values across all eight observations. This outcome was expected, as tests on simulated data had already indicated a lack of constraints in the model. This inconsistency was evidenced by the low  R$^2$  values recorded during training. As anticipated, applying the model to real observational data resulted in highly variable spin values ranging from 0.7 to -0.9 and  $R_{\text{high}}$  values spanning from 1 to 120. Given these limitations, we chose not to include results from MAD-trained static models.

\begin{table*}
  \caption{Mean outcomes from static models trained on SANE data were applied to actual data, utilizing five different train/split configurations. "Hi" and "Lo" denote the high and low frequency bands, specifically 227 GHz and 229 GHz. The standard deviation (STD) for the parameter $a_{*}$ without noise is $0.17$, 
  which becomes $0.11$ 
  when incorporating statistical errors. For $R_{\text{high}}$, the STD is $3.67$ 
  in the absence of noise and $5.29$ 
  when accounting for statistical errors.}

\label{tab:table_GRMHD_seq_for_real_obs}
  \centering
  \setlength{\tabcolsep}{5pt}
  \begin{tabular}{llllllll}
    \toprule
    & & \multicolumn{2}{c}{Without Noise} & \multicolumn{2}{c}{With Statistical Errors} \\
    \cmidrule(lr){3-4} \cmidrule(lr){5-6}
    Real Data 2017 & Parameter(s) & Hi & Lo & Hi & Lo \\
    \midrule
    April 5th & $a_{*}$ & -0.4291 & -0.6294 & -0.6877 & -0.6710 \\
              & $R{\text{high}}$ & 1.0962 & 1.8945 & 1.0399 & 2.5898 \\
    \hline
    April 6th & $a_{*}$ & -0.7845 & -0.8501 & -0.7181 & -0.8887 \\
              & $R{\text{high}}$ & 1.6892 & 4.2501 & 1.7047 & 3.0496 \\
    \hline
    April 10th & $a_{*}$ & -0.8945 & -0.8497 & -0.6359 & -0.9019 \\
               & $R{\text{high}}$ & 1.8008 & 5.3486 & 1.0709 & 16.5867 \\
    \hline
    April 11th & $a_{*}$ & -0.8920 & -0.9136 & -0.6873 & -0.8688 \\
               & $R_{\text{high}}$ & 1.0102 & 11.8475 & 1.3573 & 1.5934 \\
    \bottomrule
  \end{tabular}
  
  \label{tab:table_GRMHD_seq_for_realobs}
\end{table*}

\smallskip
Table~\ref{tab:table_GRMHD_seq_for_real_obs} summarizes the inferred spin parameter \( a_* \) from real M87* observational data. Across four observation days in 2017, with high (Hi) and low (Lo) bands analyzed separately, the predicted spin values—without noise—consistently indicated a retrograde spin, with a mean of \( \mu_{a_*} = -0.78 \) 
and a standard deviation of \( \sigma_{a_*} = 0.17 \).
When statistical errors were introduced, the predictions became more robust, with reduced variability 
\( \sigma_{a_*} = 0.11 \) 
and a slightly lower mean spin value \( \mu_{a_*} = -0.76 \).  

\smallskip

The parameter \( R_{\text{high}} \) exhibited significant variability in predictions across observations, both with and without noise.
Without noise, the predicted mean \( R_{\text{high}} \) was 
3.62 
with a standard deviation of 
3.67, 
indicating instability between observations. When statistical errors were introduced, the mean decreased slightly to 
3.62, 
while the standard deviation increased to 
5.29, 
highlighting greater instability due to noise.

Overall, the high variability in \( R_{\text{high}} \) predictions, particularly under noisy conditions, suggests that the models struggled to generalize this parameter effectively from the training simulations to real observational data.

\section{Discussion}

This study presents a method for extracting physical parameters of black holes directly from the u-v visibility spectrum using deep learning models. We trained and evaluated models on both SANE and MAD GRMHD simulation data, exploring static and sequential architectures. Our results demonstrate the feasibility of using deep learning to infer black hole properties, but also reveal challenges when applying models to real Event Horizon Telescope (EHT) observations.

One of the key observations in our study is the instability in $a_{*}$ and $R_{\text{high}}$ predictions, particularly for real data and models trained on MAD simulations. Several factors may contribute to this instability:
\begin{itemize}
    \item \textbf{Model limitations:} While our architecture performed well on SANE data, MAD-trained models showed lower $R^2$ values. However, extensive testing with alternative architectures did not yield improvements, suggesting that model design is not the primary issue.
    \item \textbf{Insufficient training data:} If data scarcity were the cause, we would expect signs of overfitting, which were not observed.
    \item \textbf{Lack of discriminating features in static MAD data:} This remains the most plausible explanation. MAD simulations may contain physical complexities that are not well captured by static representations, making parameter estimation more difficult.
\end{itemize}

In contrast, models trained on SANE data produced \textbf{stable and reliable predictions} across training, validation, and real EHT data. The static-SANE model estimated the black hole spin as $a_{\star}=-0.76 \pm  0.11$ 
and the disk property as $R_{\text{high}}=3.62 \pm 3.67$. These values were consistent across different datasets, reinforcing the robustness of the SANE-trained models. The better performance of SANE models suggests that their accretion dynamics exhibit clearer observational signatures that deep learning models can effectively capture.

The performance gap between MAD and SANE-trained models aligns with prior studies, which have noted that MAD states introduce additional turbulence and variability in jet formation, affecting observables in ways that static models struggle to capture. Additionally, EHT reconstructions are known to introduce systematic uncertainties, further complicating direct comparisons with simulations.

\subsection{Attribution Study}\label{attribution_study}
In this study, we analyze how different pairs of stations contribute to estimating the black hole spin parameter, providing insights into the interpretability of the models. Similar to how saliency maps identify the most influential pixels in image-based deep learning models \cite{simonyan2014deep, bach2015pixelwise, zeiler2014visualizing}, we apply a systematic attribution approach to evaluate the relative importance of station pair group in our dataset.

Our experiment is based on the SANE 230 GHz dataset without noise and focuses on understanding the influence of various subsets of the dataset on the predictions made by a trained neural network. To achieve this, we perform a group-masking analysis, where individual station groups are selectively masked. The resulting degradation in prediction accuracy is quantified by the change in the mean squared error (MSE), providing a measure of each group’s significance. This allows us to identify key observational baselines that contribute most to black hole spin inference.

Our dataset consists of 3,535 samples with 274 features, partitioned into 11 distinct station pair clusters. These clusters represent unique combinations of EHT baselines, including key station pairs such as AA-PV, JC-SM, and AZ-LM. Each cluster encapsulates a subset of the u-v coverage, effectively defining the spatial information available to the model.

To assess the contribution of each group, we begin by establishing a baseline prediction using all stations. We then iteratively mask each group and measure the MSE of the resulting predictions. A larger increase in MSE when a group is masked indicates a stronger reliance on that data subset, highlighting its importance in the inference process.

The process follows a structured pipeline consisting of data loading, feature extraction, and evaluation, which is detailed in Appendix \ref{appendixB}.

\subsubsection{Attribution Insights: Group Impact on Spin Estimation}

To interpret these results, we visualize the attribution effects using a key representation.

\begin{figure*}
    \centering
    \includegraphics[width=1.0\textwidth]{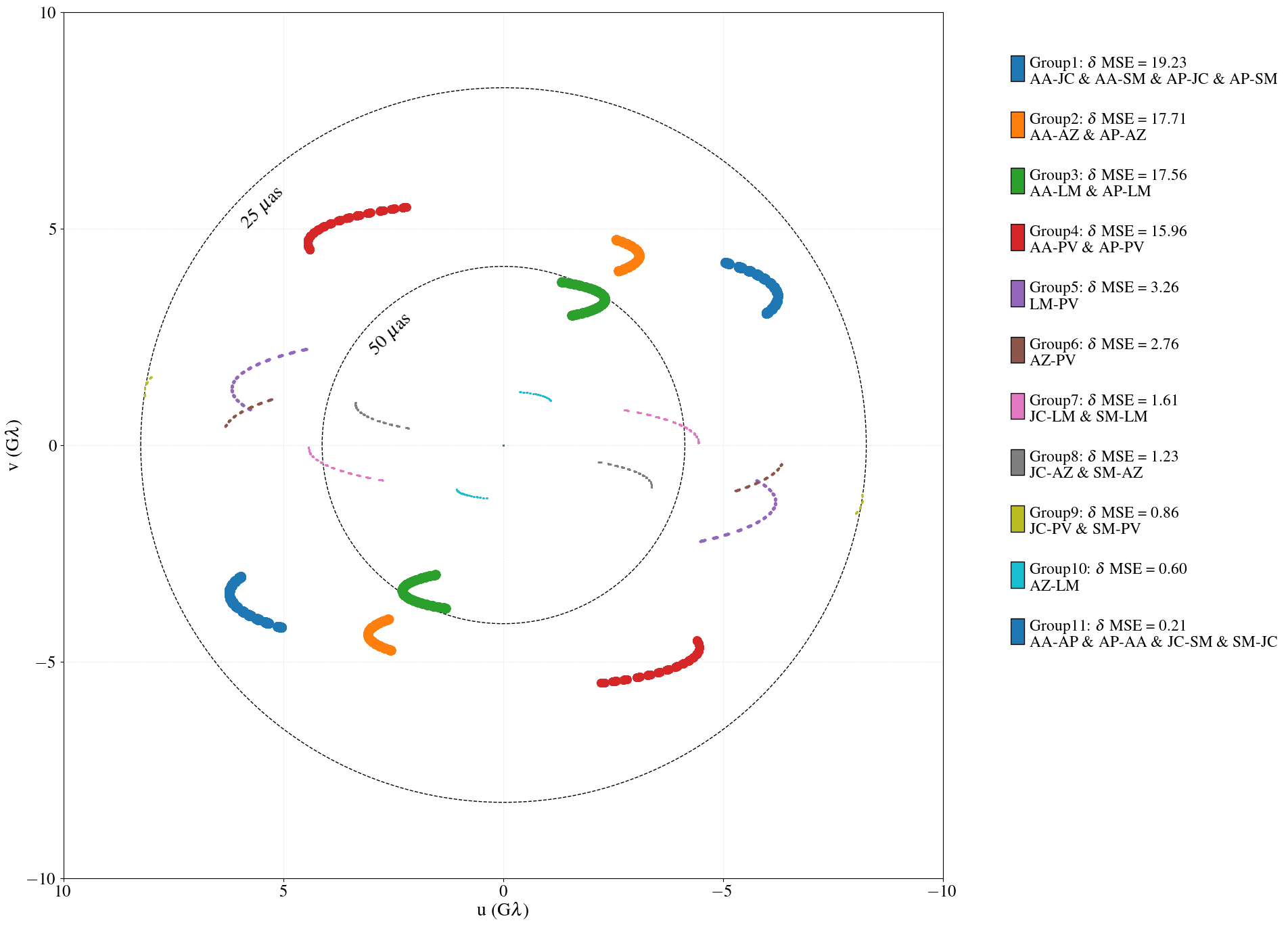}
    \caption{U-V Attribution. The size of the markers corresponds to the $\delta$ MSE values when each group is occulted. The colors indicate the different groups, as shown in the legend to the right.
    }
    \label{fig:2d_plot_coverage_with_delta_MSE}
\end{figure*}

The 2D Spatial Attribution Map (Figure~\ref{fig:2d_plot_coverage_with_delta_MSE}) represents the u-v plane coverage, where the size of the markets correlates with the change in MSE, providing a spatial view of cluster importance, and the color identifies the cluster groups.

The results reveal substantial variations in predictive significance across different station pair groups:

\begin{itemize}
    \item \textbf{High-Impact Clusters:} 
 Removing these clusters results in a substantial increase in MSE, highlighting their strong influence on spin parameter estimation. The highest impact is observed in Group 1 (AA-JC, JC-AA, AP-JC, JC-AP, AA-SM, SM-AA, AP-SM, SM-AP) with MSE = 19.25.

\item \textbf{Moderate-Impact Clusters:}
These clusters contribute meaningfully but are not as dominant. Groups such as Group 5 (LM-PV, PV-LM), Group 6 (AZ-PV, PV-AZ), Group 7 (JC-LM, LM-JC, SM-LM, LM-SM) and Group 8 (AZ-JC, JC-AZ, AZ-SM, SM-AZ) exhibit MSEs around 3.28–1.25, suggesting they contain supplementary but non-essential information.
\item \textbf{Low-Impact Clusters:}
Some station pairs appear less critical for spin inference, as masking them results in only minor changes to MSE. For example, Group 10 (AZ-LM, LM-AZ) yields an MSE of 0.23, indicating limited predictive utility.
\end{itemize}

The highest-impact groups are all baselines, including ALMA, whose very low system temperature results in the most precise measurements.  These are also long baselines, which are sensitive to emission on smaller resolved scales.  We are reassured that baselines at the origin are the least informative.  Zero baselines encode the total flux density as well as the spatial offsets via the amplitude and phase gradient, respectively, both of which have been controlled for in this study.

\subsubsection{Implications for EHT Observations}
Our findings suggest that certain station pairs encode disproportionately high information about black hole spin, likely due to their ability to capture key visibility amplitudes and phases. In contrast, lower-impact clusters contribute less unique information, potentially due to redundancy in baseline coverage or limited sensitivity to relevant spin features. This attribution study provides a data-driven perspective for optimizing future EHT array configurations, guiding the prioritization of station baselines in observational campaigns.

However, it is important to note that in this study, we did not control for signal-to-noise ratios, which could significantly affect the results and should be considered in future work.

These insights help refine our understanding of the impact of different structures in the dataset, guiding future improvements in modeling techniques and observations.

\subsection{Limitations and Future Improvements}

To ensure our models are adaptable to realistic observing conditions, we trained them on two types of simulation data: (1) raw simulations and (2) simulations with statistical errors. This approach aimed to mimic the constraints of the VLBI array, but limitations remain.

A key limitation of our current framework is the exclusive use of unpolarized MAD data. Future improvements will incorporate polarized MAD data to enhance model accuracy when applied to real EHT observations. We expect this will refine parameter estimates and provide deeper insights into black hole physics.

To further address observed limitations and variability in parameter estimation, we propose the following future improvements:
\begin{itemize}
    \item \textbf{Sequential model architectures:} Attention-based architectures may better capture temporal coherence in u-v visibility data, reducing inconsistencies across observation days.
    \item \textbf{Advanced noise augmentation:} More sophisticated noise modeling, incorporating VLBI-specific systematics, could improve robustness without sacrificing accuracy.
    \item \textbf{Refinement of $R_{\text{high}}$ predictions:} Beyond incorporating polarized MAD datasets, we will focus on improving $R_{\text{high}}$ estimates through targeted model tuning and higher-fidelity training data.
\end{itemize}

These enhancements will ensure our method remains robust under diverse observing conditions and can be extended to more complex astrophysical scenarios.

\section{Conclusion}

Deep learning models are playing an increasingly important role in astrophysics, particularly in high-resolution imaging and parameter estimation for black holes. In this study, we leveraged regressive models trained on SANE and MAD GRMHD simulations to estimate the spin ($a_{*}$) and disk properties ($R_{\text{high}}$) of M87$^{\star}$ from u-v visibility data. Our findings suggest that while SANE-based models produced stable and reliable predictions, MAD-trained models exhibited significant instability.

We evaluated both static and sequential architectures under various noise conditions. Static-SANE and sequential models successfully extracted black hole spin and disk parameters from simulated data, whereas static-MAD models struggled, highlighting the challenges of capturing MAD-state physics using static approaches. When applied to real EHT observations of M87$^{\star}$, the static-SANE model produced reasonable spin estimates ($a_{\star} = -0.76 \pm 0.10$),
reinforcing its robustness. However, disk property predictions remained uncertain ($R_{\text{high}} = 3.62 \pm 3.67$), and noise augmentation had a limited impact on improving performance.

Our results indicate that while SANE simulations provide a viable training set for deep learning-based parameter estimation, they may not fully encapsulate the complexities of M87$^{\star}$’s accretion dynamics. The poor performance of MAD-trained models suggests that additional physics—such as polarization—may be necessary to achieve accurate parameter recovery in MAD states. Future work will focus on incorporating polarized MAD simulations into the training pipeline. This approach is expected to enhance model precision and yield deeper insights into black hole accretion physics.

\section*{Acknowledgments}
\noindent
This project/publication is funded in part by the Gordon and Betty Moore Foundation (Grant \#8273.01). It was also made possible through the support of a grant from the John Templeton Foundation (Grant \#62286).  The opinions expressed in this publication are those of the author(s) and do not necessarily reflect the views of these Foundations.
CG thanks the AstroAI team for helpful discussions. CG was supported by AstroAI at the Center for Astrophysics | Harvard \& Smithsonian.

\section*{Data Availability}
GRMHD simulations and real observations from the Event Horizon Telescope (EHT) were used in this study. 
GRMHD simulation data are available at \citet{Roelofs2021} and \citet{Fromm2022}, including detailed information about the simulations. 
EHT observations are also accessible at \url{eventhorizontelescope.org/for-astronomers/data}. 
All these data was processed using the \texttt{ehtim} scripts, accessible at \url{github.com/achael/eht-imaging}.

\bibliographystyle{mnras}
\bibliography{manuscript} %

\newpage
\appendix

\section{Architectures and Model training}
\label{appendixA}
\subsection{Static Pipeline}
\subsubsection{Autoencoder Training}
\label{Autoencoder_Training_Static}

The autoencoder architecture is described in Table \ref{tab:autoencoder_architecture}.
The model was trained to minimize the reconstruction loss of the input data.
The training data was split into 80\% for training and 20\% for validation, with a further 50/50 split of the validation set for testing.

The model was compiled with the \texttt{Adam} optimizer, using an exponentially decaying learning rate starting at $10^{-4}$. The loss function used was the mean squared error (MSE). Training was performed using the \texttt{fit} function in TensorFlow:

\begin{itemize}
    \item \textbf{Training data:} The autoencoder was trained on the input data with the target being the last feature (indexed as $[:, :, 2:]$).
    \item \textbf{Epochs:} The maximum number of epochs was set to 3000.
    \item \textbf{Batch size:} A batch size of 64 was chosen to balance training speed and memory usage.
    \item \textbf{Callbacks:} Early stopping was employed to monitor validation loss and halt training after 300 epochs without improvement.
\end{itemize}

Validation loss and training loss were logged
The trained autoencoder achieved a reconstruction MSE of $1e^{-5}$ on the test set.

\subsubsection{CNN Training}

The encoder component of the trained autoencoder was used to extract feature embeddings from the input data. 

These embeddings, with a size of 50 (as described in Table \ref{tab:autoencoder_architecture}), were used as inputs to the CNN model.

The CNN architecture is described in Table \ref{tab:cnn_architecture_detailed}. The model was trained as follows:

\begin{itemize}
    \item \textbf{Training data:} The embedding vectors were reshaped into $(N, 1, 50)$ for compatibility with the 1D CNN layers.
    \item \textbf{Loss functions:} The loss function was mean squared error (MSE) for both outputs. A log transformation was applied to the \texttt{Rhigh} labels to normalize the values before training.
    \item \textbf{Optimizer:} The same exponentially decaying learning rate schedule as the autoencoder was used.
    \item \textbf{Epochs:} Training was conducted for a maximum of 2000 epochs with early stopping, similar to the autoencoder.
    \item \textbf{Batch size:} A batch size of 64 was used.
\end{itemize}

The \texttt{fit} function was configured with two separate outputs (\texttt{spin} and \texttt{Rhigh}) for the two regression branches of the CNN.

\subsubsection{Evaluation}

The trained models were evaluated on both the test set and real observational data. The autoencoder's reconstruction performance was measured using the MSE between the original and reconstructed data, while the CNN model was evaluated using the $R^2$ metric for both output branches:

Additionally, the real observational data were processed using the encoder and then passed to the CNN model for predictions.

\subsection{Sequencial Pipeline}
\subsubsection{Autoencoder Training}

The autoencoder model for the sequential pipeline, the architecture and the training was exactly the same as the one used for the static pipeline as detailed in \ref{Autoencoder_Training_Static}.

\subsubsection{LM Model Training}
\label{LM_Training_Sequential}

A BiLSTM-based language model was trained to learn sequential representations from the encoded data. The encoder component of the trained autoencoder was used to extract feature embeddings from the input data. These embeddings, with a size of 50 (as described in Table \ref{tab:autoencoder_architecture}), were used as inputs to the LM model. 
In between, the data processing pipeline involved segmenting the encoded data into fixed-length sequences and normalizing the values that we used as the LM input, before training.

With this prepared segmented data, the model was trained to predict the next step in the sequence using a shifted version of the training data.

The LM model is described in Table \ref{tab:lstm_model_architecture}

The training configuration was as follows:

\begin{itemize}
    \item \textbf{Loss function:} Mean squared error (MSE) between predicted and actual next-step embeddings.
    \item \textbf{Optimizer:} \texttt{Adam} with an exponentially decaying learning rate (starting at $10^{-3}$, decaying by 0.9 every 1000 steps).
    \item \textbf{Batch size:} 64.
    \item \textbf{Epochs:} Maximum of 2000 epochs with early stopping (patience = 300).
    \item \textbf{Validation:} 20\% of the training set was used for validation.
\end{itemize}

\subsubsection{BiLSTM Model Training}

A language model (LM) \ref{LM_Training_Sequential} was first trained on the sequence data to learn meaningful representations before training the full BiLSTM model.

The full BiLSTM model was trained using the learned sequential representations from the LM model. For that, the LM model's weights were then transferred to initialize the BiLSTM network.
This model was trained to predict the \texttt{spin} and \texttt{Rhigh} labels.

The BiLSTM architecture is described in Table \ref{tab:full_model_architecture}. The model was trained as follows:

\begin{itemize}
    \item \textbf{Training data:} The segmented input sequences were formatted into $(N, T, 50)$ for compatibility with the BiLSTM layers.
    \item \textbf{Loss function:} The mean squared error (MSE) loss was used for both \texttt{spin} and \texttt{Rhigh} predictions. A log transformation was applied to \texttt{Rhigh} labels to improve numerical stability.
    \item \textbf{Optimizer:} The optimizer followed the same exponentially decaying learning rate schedule as the autoencoder.
    \item \textbf{Epochs:} Training was conducted for a maximum of 2000 epochs with early stopping.
    \item \textbf{Batch size:} A batch size of 64 was used.
    \item \textbf{Pretraining:} The LM model was trained on the segmented sequences before initializing the BiLSTM model.
\end{itemize}

The \texttt{fit} function was configured with two separate outputs for the two regression branches of the BiLSTM model.

\subsubsection{Evaluation}

The trained models were evaluated on the test set. The autoencoder's reconstruction performance was measured using the MSE between the original and reconstructed data, while the BiLSTM model was evaluated using the $R^2$ metric for both output branches.

\section{Attribution Study}
\label{appendixB}
The process follows a structured pipeline consisting of data loading, feature extraction, clustering, and evaluation. Below, we describe each step in detail.

\subsection{Data Preparation}

The dataset is loaded from a precomputed NumPy array, which contains the processed observations. Each sample in the dataset has a shape of $(274,4)$, representing 274 temporal observations with four extracted features. We conducted the study with the 3535 frames available, made up of five different label groups ($-0.94, -0.50, 0.00, 0.50, 0.94$), of 707 samples each.

\subsection{Neural Network Architectures}

To extract meaningful features from the data, we employ an encoder network and a CNN model. The encoder transforms the raw input into a lower-dimensional latent representation, while the CNN model refines these embeddings for final predictions. The specific architectures used for both models are presented in Table \ref{tab:autoencoder_architecture} for the encoder and \ref{tab:cnn_architecture_detailed} for the 1D CNN.

\subsection{Embedding Extraction and Normalization}

Once the data is loaded, we use the trained encoder model to extract latent representations for each sample:

\begin{equation}
    E = \text{Encoder}(X)
\end{equation}

where $X$ represents the input data, and $E$ is the corresponding embedding. To ensure comparability, these embeddings are normalized using a \texttt{StandardScaler}, which applies:

\begin{equation}
    E' = \frac{E - \mu}{\sigma}
\end{equation}

where $\mu$ and $\sigma$ represent the mean and standard deviation computed across the entire dataset.

\subsection{Clustering of Stations}

The core of the experiment involves clustering the measurements based on their position in the (u,v)-plane.  Specifically, we group together measurements by their baselines, where co-located stations (like ALMA and APEX) are considered equivalent.
This presents a total of 11 groups described in table \ref{tab:group_pairs}.  This allows us to determine which baselines are most important for parameter inference in our study.

The initials of those stations correspond to:

\begin{itemize}
    \item \textbf{AA}: Atacama Large Millimeter/submillimeter Array (ALMA), Chile
    \item \textbf{AP}: Atacama Pathfinder Experiment (APEX), Chile
    \item \textbf{AZ}: Arizona Radio Observatory Submillimeter Telescope (ARO-SMT), USA
    \item \textbf{JC}: James Clerk Maxwell Telescope (JCMT), Hawaii, USA
    \item \textbf{SM}: Submillimeter Array (SMA), Hawaii, USA
    \item \textbf{LM}: Large Millimeter Telescope Alfonso Serrano (LMT), Mexico
    \item \textbf{PV}: Institut de Radioastronomie Millimétrique (IRAM) 30m Telescope, Pico Veleta, Spain
\end{itemize}

\subsection{Prediction and Error Analysis}

After determining the clusters to use, we further analyze the predictive performance. Each cluster's data is passed through the encoder and subsequently the CNN model:

\begin{equation}
    \hat{Y} = \text{CNN}(\text{Encoder}(\mathcal{C}_i))
\end{equation}

where $\hat{Y}$ represents the predicted label.

The error is quantified using the Mean Squared Error (MSE):

\begin{equation}
    \text{MSE} = \frac{1}{N} \sum_{i=1}^{N} (Y_i - \hat{Y}_i)^2
\end{equation}

where $Y_i$ denotes the ground truth label.

To obtain the $\delta$ MSE we subtract the MSE minus the $\hat{MSE}$ baseline where all the clusters are present:

\begin{equation}
    \text{$\delta$ MSE} = MSE - \hat{MSE}
\end{equation}

\subsection{Visualization and Interpretation}

\begin{figure*}
    \centering
    \includegraphics[width=1.0\textwidth]{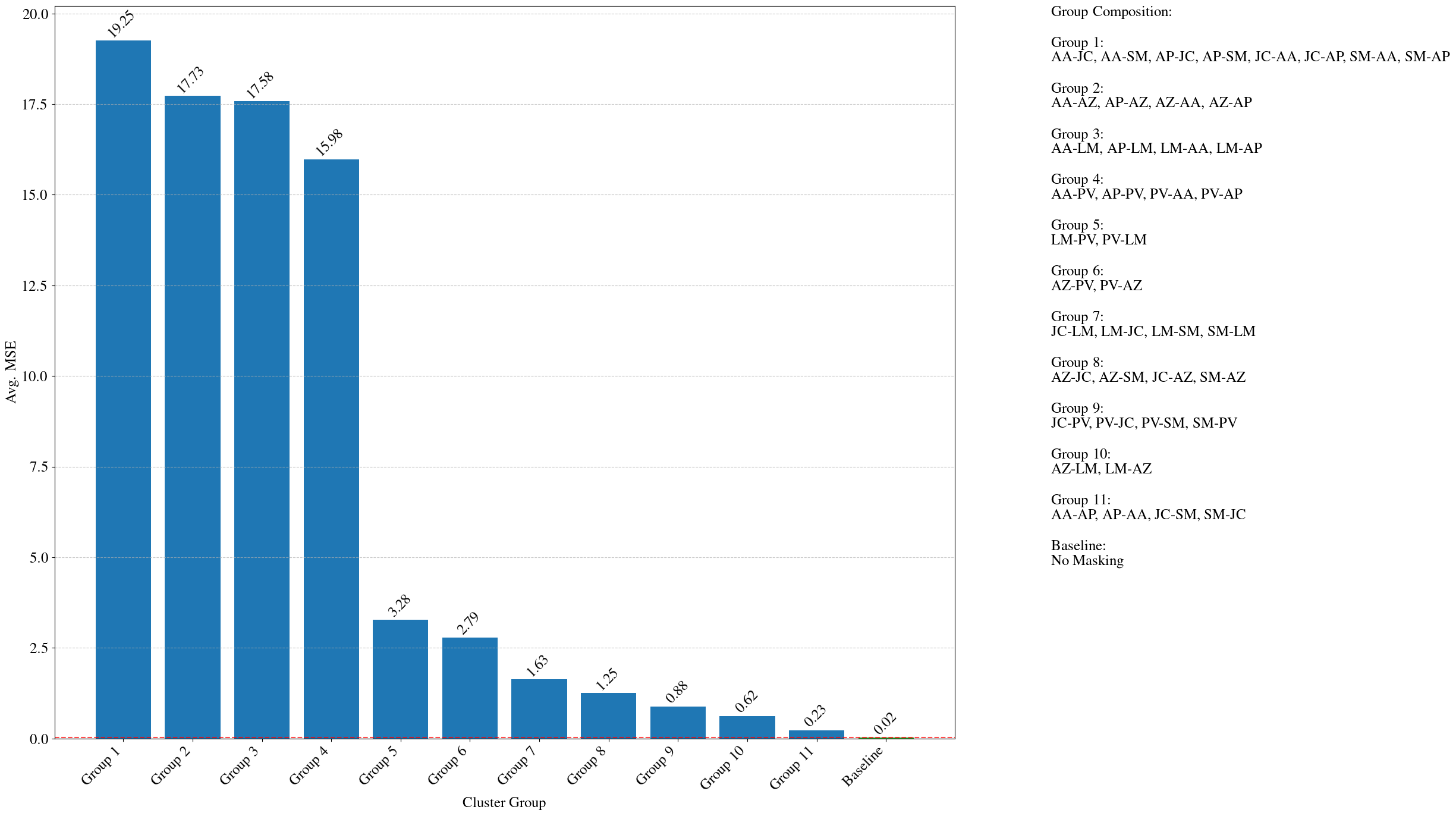}
    \caption{Cluster Importance bar chart based on MSE with the MSE baseline for comparison}
    \label{fig:baseline_stations_comparison_no_colormap}
\end{figure*}

Figure~\ref{fig:baseline_stations_comparison_no_colormap} compares each arc MSE to the $\hat{MSE}$ baseline.

Please refer to section \ref{attribution_study} and Figure~\ref{fig:2d_plot_coverage_with_delta_MSE} for further details about the interpretation and visualization of the attribution map.

\begin{table*}
\centering
\caption{Autoencoder Model Architecture}
\label{tab:autoencoder_architecture}
\begin{tabular}{lcc}
\toprule
\textbf{Layer} & \textbf{Type} & \textbf{Output Channels} \\
\midrule
\multicolumn{3}{c}{\textbf{Encoder}} \\
\midrule
Input & Input Layer & 4 \\
Conv1 & Conv1D (filters=10, kernel=3, ELU) & 10 \\
Conv2 & Conv1D (filters=20, kernel=3, ELU) & 20 \\
Conv3 & Conv1D (filters=50, kernel=3, ELU) & 50 \\
Conv4 & Conv1D (filters=50, kernel=3, ELU) & 50 \\
Conv5 & Conv1D (filters=50, kernel=3, ELU) & 50 \\
Output & GlobalAvgPool1D & 50 \\
\midrule
\multicolumn{3}{c}{\textbf{Decoder}} \\
\midrule
Conv1 & Conv1DTranspose (filters=50, kernel=3, ELU) & 50 \\
Conv2 & Conv1DTranspose (filters=50, kernel=3, ELU) & 50 \\
Conv3 & Conv1DTranspose (filters=50, kernel=3, ELU) & 50 \\
Conv4 & Conv1DTranspose (filters=20, kernel=3, ELU) & 20 \\
Conv5 & Conv1DTranspose (filters=10, kernel=3, ELU) & 10 \\
Output & Conv1DTranspose (filters=2, kernel=3, linear) & 2 \\
\bottomrule
\end{tabular}
\end{table*}

\begin{table*}
\centering
\caption{LSTM Model Architecture}
\label{tab:lstm_model_architecture}
\begin{tabular}{llc}
\toprule
\textbf{Layer} & \textbf{Type} & \textbf{Output Channels} \\
\midrule
Input & Input Layer & 50 \\
LSTM & Bidirectional LSTM (units=64, kernel\_regularizer=l2(0.001), return\_sequences=True) & 128 \\
Dropout & Dropout (rate=0.3) & 128 \\
LSTM & Bidirectional LSTM (units=64, kernel\_regularizer=l2(0.001), return\_sequences=True) & 128 \\
Dropout & Dropout (rate=0.3) & 128 \\
Output & Dense (units=50, activation=linear) & 50 \\
\bottomrule
\end{tabular}
\end{table*}

\begin{table*}
\centering
\caption{Sequential Bi-LSTM, Multitask Model Architecture}
\label{tab:full_model_architecture}
\begin{tabular}{llc}
\toprule
\textbf{Layer} & \textbf{Type} & \textbf{Output Channels} \\
\midrule
\textbf{LSTM Layers} & & \\
\midrule
Inputs & Input Layer & 50 \\
LSTM1 & Bidirectional LSTM (units=64, kernel\_regularizer=l2(0.001), return\_sequences=True) & 128 \\
Dropout1 & Dropout (rate=0.3) & 128 \\
LSTM2 & Bidirectional LSTM (units=64, kernel\_regularizer=l2(0.001), return\_sequences=False) & 128 \\
Dropout2 & Dropout (rate=0.3) & 128 \\
\midrule
\textbf{Spin Output Branch} & & \\
\midrule
Regression1 & Dense (units=128, activation=tanh, kernel\_regularizer=l2(0.001)) & 28 \\
Regression2 & Dense (units=64, activation=tanh, kernel\_regularizer=l2(0.001)) & 64 \\
Regression3 & Dense (units=32, activation=tanh, kernel\_regularizer=l2(0.001)) & 32 \\
Output & Dense (units=1, activation=tanh) & 1 \\
\midrule
\textbf{Rhigh Output Branch} & & \\
\midrule
Regression1 & Dense (units=128, activation=relu, kernel\_regularizer=l2(0.001)) & 128 \\
Regression2 & Dense (units=64, activation=relu, kernel\_regularizer=l2(0.001)) & 64 \\
Regression3 & Dense (units=32, activation=relu, kernel\_regularizer=l2(0.001)) & 32 \\
Output & Dense (units=1, activation=relu) & 1 \\
\bottomrule
\end{tabular}
\end{table*}

\begin{table*}
\centering
\caption{Static 1D CNN, Multitask Model Architecture}
\label{tab:cnn_architecture_detailed}
\begin{tabular}{llc}
\toprule
\textbf{Layer} & \textbf{Type} & \textbf{Output Channels} \\
\midrule
Inputs & Input Layer & 1, 50 \\
Conv1 & Conv1D (filters=32, kernel size=3, activation=ReLU, padding=same) & 1, 32 \\
BatchNormalization1 & BatchNormalization & 1, 32 \\
Conv2 & Conv1D (filters=64, kernel size=3, activation=ReLU, padding=same) & 1, 64 \\
BatchNormalization2 & BatchNormalization & 1, 64 \\
Conv3 & Conv1D (filters=128, kernel size=3, activation=ReLU, padding=same) & 1, 128 \\
BatchNormalization3 & BatchNormalization & 1, 128 \\
Flatten & Flatten & 128 \\
\midrule
\textbf{Spin Regression Branch} & & \\
\midrule
Dense1 & Dense (units=64, activation=ReLU) & 64 \\
Dropout & Dropout (rate=0.4) & 64 \\
Dense2 & Dense (units=128, activation=ReLU) & 128 \\
Dense3 & Dense (units=64, activation=ReLU) & 64 \\
Output & Dense (units=1, activation=tanh) & 1 \\
\midrule
\textbf{$R_{\text{high}}$ Regression Branch} & & \\
\midrule
Dense1 & Dense (units=64, activation=ReLU) & 64 \\
Dropout & Dropout (rate=0.4) & 64 \\
Dense2 & Dense (units=128, activation=ReLU) & 128 \\
Dense3 & Dense (units=64, activation=ReLU) & 64 \\
Output & Dense (units=1, activation=linear) & 1 \\
\bottomrule
\end{tabular}
\end{table*}

\begin{table}
    \centering
    \caption{Group pairings}
    \begin{tabular}{ll}
        \textbf{Group} & \textbf{Pairs} \\
        \hline
        Group 1  & \{AA-JC, JC-AA, AP-JC, JC-AP, AA-SM, SM-AA, AP-SM, SM-AP\} \\
        Group 2  & \{AA-AZ, AZ-AA, AP-AZ, AZ-AP\} \\
        Group 3  & \{AA-LM, LM-AA, AP-LM, LM-AP\} \\
        Group 4  & \{AA-PV, PV-AA, AP-PV, PV-AP\} \\
        Group 5 & \{LM-PV, PV-LM\} \\
        Group 6 & \{AZ-PV, PV-AZ\} \\
        Group 7  & \{JC-LM, LM-JC, SM-LM, LM-SM\} \\
        Group 8  & \{AZ-JC, JC-AZ, AZ-SM, SM-AZ\} \\
        Group 9  & \{JC-PV, PV-JC, SM-PV, PV-SM\} \\
        Group 10  & \{AZ-LM, LM-AZ\} \\
        Group 11  & \{AA-AP, AP-AA, JC-SM, SM-JC\} \\     
    \end{tabular}
    \label{tab:group_pairs}
\end{table}

\bsp	%
\label{lastpage}
\end{document}